\documentclass[aps,preprint,nofootinbib,showpacs,11pt]{revtex4}
\usepackage{amsfonts,amsthm}
\usepackage{amssymb,amsmath}
\usepackage{graphicx}

\usepackage{color}

\begin{document}
\title{A brief history of gravitational wave research}

\author{Chiang-Mei Chen$^{1}$} \email{cmchen@phy.ncu.edu.tw}
\author{James M. Nester$^{1,2,3}$} \email{nester@phy.ncu.edu.tw}
\author{Wei-Tou Ni$^{4,5}$} \email{weitou@gmail.com}

\affiliation{$^1$Department of Physics, National Central University, Chungli 32001, Taiwan, ROC}
\affiliation{$^2$Graduate Institute of Astronomy, National Central University, Chungli 32001, Taiwan, ROC}
\affiliation{$^3$Leung Center for Cosmology and Particle Astrophysics, National Taiwan University, Taipei 10617, Taiwan, ROC}
\affiliation{$^4$Department of Physics, National Tsing Hua University, Hsinchu 30013, Taiwan, ROC}
\affiliation{$^5$School of Optical-Electrical and Computer Engineering, University of Shanghai for Science and Technology, Shanghai 200093, China}



\begin{abstract} 
For the benefit of the readers of this journal, the editors requested that we prepare a brief review of the history of the development of the theory, the experimental attempts to detect them, and the recent direct observations of gravitational waves (GWs).  The theoretical ideas and disputes beginning with Einstein in 1916 regarding the existence and nature of gravitational waves and the extent to which one can rely on the electromagnetic analogy, especially the controversies regarding the quadrupole formula and whether gravitational waves carry energy, are discussed.  The  theoretical conclusions eventually received strong observational support from the binary pulsar. This provided compelling, although indirect, evidence for gravitational waves carrying away energy---as predicted by the quadrupole formula.
On the direct detection experimental side, Joseph Weber started more than fifty years ago. In 1966, his bar for GW detection reached a strain sensitivity of a few times $10^{-16}$. His announcement of coincident signals (now considered spurious), stimulated many experimental efforts from room temperature resonant masses to cryogenic detectors and laser-interferometers. Now there are km-sized interferometric detectors (LIGO Hanford, LIGO Livingston, Virgo and KAGRA). Advanced LIGO first reached a strain sensitivity of the order of $10^{-22}$. During their first 130 days of observation (O1 run), with the aid of templates generated by numerical relativity, they did make the first detections:  two 5-$\sigma$ GW events and one likely event.
Besides earth-based GW detectors, the drag-free sensitivity of the LISA Pathfinder has already reached to the LISA goal level, paving the road for space GW detectors. Over the whole GW spectrum (from aHz to THz) there are efforts for detection, notably the very-low-frequency band (pulsar timing array [PTA], 300 pHz -- 100 nHz) and the extremely-low (Hubble)-frequency (cosmic microwave background [CMB] experiment, 1 aHz -- 10 fHz).
\end{abstract}

\maketitle

\section{Introduction}
One of the most exciting recent results in physics has been, after much effort, the direct detection of gravitational waves (GWs) very nearly a century after being predicted by Einstein based on his theory of gravity, general relativity (GR).  Here, for the benefit of readers of this journal, we offer a brief review of gravitational waves: the theoretical history, the experiments and the recent observations.  Our plan is to first mention some forerunners, then Einstein's work of 1916 and 1918 and that of Eddington in 1922, 23 along with a short review of gravitational wave theory.  We then give an account of the remarkable Einstein-Rosen paper of 1936.  This is followed by some discussion of the theoretical disputes regarding gravitational waves, whether they could transport energy, and the quadrupole formula.  For this material we largely rely on the book \textit{Traveling at the speed of thought} by Daniel Kennefick~\cite{Kennefick}.  We hope that our brief overview of some of the highlights of that work will encourage those interested to read all the details in his fascinating work.  Kennefick insightfully discusses the aforementioned disputes as being between skeptics and non-skeptics regarding the extent that one can rely on the electromagnetic analogy when considering gravitational radiation.  Following our theoretical discussion, we then turn to the experiments and observations.  On the one hand the decay of the orbit of the binary pulsar has been regarded as strong support for gravitational waves carrying away energy as predicted by the quadrupole formula~\cite{Kennefick:2014bba}.  On the other hand efforts were made to detect gravitational waves directly (for a detailed history of these efforts see Harry Collins \textit{Gravity's Shadow}~\cite{Collins04}).  Joseph Weber's pioneering efforts using resonant masses stimulated other efforts.  Eventually people turned to laser interferometry. Numerical simulations of the expected waveforms has played an important role.   Early this year the LIGO-Virgo collaboration announced the detection of GW150914~\cite{Abbott:2016blz}; in June of this year (which is the 100th anniversary of Einstein's first paper on gravitational waves~\cite{AE16}) the collaboration presented its summary of the first Advanced LIGO observing run, including a second binary black hole merger event, GW151226~\cite{Abbott:2016nmj}, and a probable black hole merger event, LVT151012. Templates generated by numerical relativity played an important role in recognizing these events.

Einstein said in his 1916 paper ``This expression (for the radiation $A$) would get an additional factor $1/c^4$ if we would measure time in seconds and energy in Erg (erg). Considering $\kappa = 1.87 \times 10^{-27}$ (in units of cm and gm),\footnote{Here $\kappa = 8 \pi G_N/c^2$.}
it is obvious that $A$ has, in all imaginable cases, a practically vanishing value.'' Indeed at that time, possible expected source strengths and the detection capability had a huge gap. However, with the great strides in the advances of astronomy and astrophysics and in the development of technology, this gap has largely been closed. A white dwarf star was discovered in 1910 with its density soon estimated. Now we understand that GWs from white dwarf binaries in our Galaxy form a stochastic GW background (``confusion limit'')~\cite{Bender:1997hs} for space (low frequency) GW detection in GR. The characteristic strain for the confusion limit is about $10^{-20}$ in the 0.1 -- 1 mHz band. As to individual sources, some can have characteristic strain around this level for frequency 1 -- 3 mHz in the low-frequency band. One hundred year ago, the sensitivity of astrometric observation through the atmosphere around this band was about 1 arcsec. This means the strain sensitivity to GW detection was about $10^{-5}$; 15 orders away from the required sensitivity. The first artificial satellite Sputnik was launched in 1957. The technological demonstration mission LISA Pathfinder was launched on 3 December, 2015. It successfully demonstrated the drag-free requirement of the LISA GW space mission concept; the major issue in the technological gap of 15 orders of magnitude has been abridged during these hundred years~\cite{Armano:2016bkm}. These considerations should also be readily applicable to other GW space mission concepts~\cite{Kuroda:2015owv}. However, at present the space GW missions are only expected to be launched more than a decade later. In the LIGO discovery of 3 GW events, the maximum peak strain intensity is $10^{-21}$; the frequency range is 30 -- 450 Hz. Strain gauge in this frequency region might reach $10^{-5}$ about 100 years ago; thus, the technology gap was 16 orders of magnitudes. The Michelson interferometer for the Michelson-Morley experiment~\cite{Michelson:1887zz} had a sensitivity of strain $(\Delta l / l)$ of $5 \times 10^{-10}$ with a 0.01 fringe detectability and a 11 m path length; however, the appropriate test mass suspension system with fast (30 -- 450 Hz in the high-frequency GW band) white-light observing system was lacking. Summarizing, the detection gap between the astrophysical GW source strength and the technological achievable sensitivity was about 15 -- 16 orders of magnitude in amplitude in the low-frequency and high-frequency GW bands 100 years ago.

\section{Forerunners}
The idea of gravitational waves did not begin with Einstein and GR.  As far as we know Clifford was the first to have such a vision, he imagined \emph{curvature waves}:

\begin{quotation}
``I hold that (1) small portions of space \emph{are} in fact of a nature analogous to little hills on a surface which is on average flat; namely that the ordinary laws of geometry are not valid for them. (2) That the property of being curved or distorted is continually being passed on from one portion of space to another after the manner of a wave. \dots''~\cite{Clifford76}\footnote{Clifford's idea was even more amazing, he envisioned a kind of unified field theory where all of material reality was just propagating waves of the curvature of 3-dimensional space.}
\end{quotation}
Nowadays we identify waves of spacetime curvature with \emph{gravitational waves}.
Poincar\'e~\cite{Po1905, Katzir} may have been the first to actually use the term \emph{gravitational wave}, he speculated about (special) relativistic gravity which would have waves of acceleration, a retarded attractive force that propagated at speed $c$.

With the development of special relativity, some, naturally, considered a Lorentz covariant scalar theory. Perhaps the best known was due to Nordstr\"om in 1913~\cite{Nordstrom}; it had waves that propagated at speed $c$.  This theory can, in fact, be reexpressed in \emph{generally covariant} form, with the Riemann curvature scalar being proportional to the trace of the matter energy-momentum density along with a \emph{vanishing} Weyl conformal curvature tensor:
\begin{equation}
R = 3 \kappa T, \qquad C^\alpha{}_{\beta\mu\nu} = 0; \qquad \kappa := \frac{8 \pi G_N}{c^4},
\end{equation}
as was demonstrated by Einstein and Fokker in 1914~\cite{EinFokk14}---at a time when Einstein still very much doubted that there could be a \emph{generally covariant} gravity theory for a dynamical spacetime metric.  Nordstr\"om's theory predicts no bending of light in a gravitational field, and thus  does not satisfy Einstein's \emph{equivalence principle}.

\section{Einstein's predictions of 1916 and 1918}
On November 25, 1915 Einstein was finally able to present a generally covariant theory of gravitation, \emph{general relativity} (GR).
His field equation, now usually written in the form
\begin{equation}
G_{\mu\nu} = \kappa T_{\mu\nu},
\end{equation}
describes how the energy-momentum density of matter, $T_{\mu\nu}$, generates the curvature of spacetime.
Here the \emph{Einstein tensor} is $G_{\mu\nu} := R_{\mu\nu} - \frac12 g_{\mu\nu} R$, with $g_{\mu\nu}$ being the spacetime metric tensor (i.e., $ds^2 = g_{\mu\nu} dx^\mu dx^\nu$) and $R_{\mu\nu}$ is the Ricci curvature tensor, obtained by contracting the Riemann curvature tensor, $R_{\mu\nu} := R^\alpha{}_{\mu\alpha\nu}$, where
\begin{equation}
R^\alpha{}_{\beta\mu\nu} := \partial_\mu \Gamma^\alpha{}_{\beta\nu} - \partial_\nu \Gamma^\alpha{}_{\beta\mu} + \Gamma^\alpha{}_{\gamma\mu} \Gamma^\gamma{}_{\beta\nu} - \Gamma^\alpha{}_{\gamma\nu} \Gamma^\gamma{}_{\beta\mu},
\end{equation}
with the Christoffel connection given in terms of the metric by
\begin{equation}
\Gamma^\alpha{}_{\beta\gamma} := \frac12 g^{\alpha\delta} (\partial_\beta g_{\delta\gamma} + \partial_\gamma g_{\delta\beta} - \partial_\delta g_{\beta\gamma}).
\end{equation}

\subsection{February 1916}
As we will discuss  in more detail later, on February 11, 2016, the LIGO collaboration announced the first direct detection of gravitational waves~\cite{Abbott:2016blz}. Almost 100 years earlier, just three months after finally finding the field equations for his GR theory, Einstein wrote to Schwarzschild:


\begin{quotation}
``Since then [Nov 4], I have handled Newton's case differently according to the final theory.---Thus there are no gravitational waves analogous to light waves. This is probably also related to the one-sidedness of the sign of scalar $T$, incidently. (Nonexistence of the ``dipole''.)''~\cite{cpae} Vol.~8, Doc.~194, 19 Feb.~1916.
\end{quotation}
This was truly amazing: \emph{no gravitational waves}!  For any interaction field one naturally expects waves.
We do not really know exactly how Einstein came to this remarkable conclusion.  Kennefick makes a reasonable case that Einstein was working out to higher order the perturbation type calculation he had used in his Mercury perihelion calculation of 18 November.  The Newtonian effects are of order $G_N m/r\sim v^2$, the next order relativistic corrections vanish, the first non-vanishing ``post-Newtonian'' corrections have an additional factor of $(v/c)^2$, it seems likely that Einstein went beyond this and found vanishing corrections at order $(v/c)^3$.  However, it is now known that to see the damping due to radiation reaction from the emission of gravitational waves one must actually go to the so-called 2.5 post-Newtonian order, i.e., an additional factor of $(v/c)^5$.  Einstein may have gone far enough to see dipole radiation, had there been any.  It appears that he understood (as had already been clearly argued by Abraham) that for purely attractive gravity, unlike the case for electromagnetism with opposite charges, there could be no dipole radiation. As far as we can tell there is no evidence that he thought at this point in time beyond dipole radiation.

\subsection{June 1916}
But in a few months he changed his mind on the subject of gravitational radiation (and he will do this a couple of more times).  He was trying to do some approximate integration of his equations and was not making much progress.  As happened on other occasions, his extensive network of correspondents proved helpful.  On this occasion it was de Sitter who gave him the hint he needed---that a coordinate condition different from the one that he had been relying on ($\sqrt{|g|}=1$) would be more suitable.  It so happens that we are writing this section in June 2016, at a historical time---just after the presention of the results of the first advanced LIGO observing run~\cite{Abbott:2016blz}, and also, coincidentally, on the occasion of the centenary anniversary of the prediction of gravitational waves.  One hundred years earlier, in June 1916, Einstein published a paper entitled ``Approximate integration of the field equations of gravitation''~\cite{AE16}.  In it he developed the weak-field linearized theory of GR.  Although he made more than his usual number of mistakes (one rather serious) he nevertheless pretty much laid out the pattern of analysis still used today. He, following de Sitter's suggestion, had determined the appropriate coordinate gauge condition and predicted the existence of gravitational waves that propagate at speed $c$ generated by a time dependent source quadrupole moment.
He found 3 types of waves (longitudinal-longitudinal, longitudinal-transverse, and transverse-transverse) and argued (based, however, on incorrect expressions) that only the 3rd type carried energy, while the other two were dependent on the choice of coordinates.


\subsection{Einstein 1918}
By way of correspondence with Nordstr\"om, Einstein learned of his June 1916 mistakes.  In January 1918 he published a paper entitled ``On gravitational waves''~\cite{AE18}. This work follows the same pattern set in 1916, but with (almost all of) the mistakes corrected.  Its content is very much like what one finds in the textbooks today.  The magnitude for the quadrupole formula was corrected from $\kappa/(24\pi)$ to $\kappa/(80\pi)$, and a major error (which had marred many of the details of the earlier work) concerning the linearized energy-momentum conservation was corrected, so this time the discussion of the two ``coordinate'' wave modes that carried no energy was on a sound footing.  In the next section we will present a short overview of the basic theory of gravity waves.

\subsection{Eddington 1922,1923}
Eddington in 1922~\cite{Eddington22} reexamined the issue with skeptical eyes.  Rather than looking at the metric (which includes coordinate waves) he looked at what Clifford had envisioned almost 5 decades earlier: \emph{waves of curvature}.  This perspective verified that the coordinate waves (in modern language, ``pure gauge degrees of freedom'') could propagate at any speed (in his words at ``the speed of thought''), but they carried no energy.  Eddington basically confirmed what Einstein had obtained in 1918, although he did find a small correction: the quadrupole amplitude should be $\kappa/(40\pi)$. In 1923 Eddington calculated the quadrupole radiation emitted by a spinning rod via the radiation reaction due to retardation~\cite{Eddington23}.  As we will discuss, there will be major controversies many decades later regarding the analogous calculations for gravitating quadrupoles.



\section{A short overview of gravitational wave theory}
Here we present a short overview of gravitational wave theory, for a more complete treatment the reader is referred to textbooks, e.g.~\cite{Weber61, MTW, Maggiore}.

\subsection{Weak Gravitational Fields: Linear Gravity}
Consider the case where the spacetime metric is a small perturbation, $h_{\mu\nu} \; (|h_{\mu\nu}| \ll 1)$, of Minkowski flat spacetme, $\eta_{\mu\nu} = {\rm diag}(-1, +1, +1, +1)$, with $x^\mu = (ct, x, y, z)$:
\begin{equation}
g_{\mu\nu} = \eta_{\mu\nu} + h_{\mu\nu} = g^{(0)}_{\mu\nu} + g^{(1)}_{\mu\nu}.
\end{equation}
This leads to the following leading (first) order expressions for the Christoffel symbol, Ricci tensor and scalar curvature as
\begin{eqnarray}
\Gamma^{(1)\mu}{}_{\alpha\beta} &=& \frac12 \eta^{\mu\nu} \left( \partial_\alpha h_{\nu\beta} + \partial_\beta h_{\nu\alpha} - \partial_\nu h_{\alpha\beta} \right),
\\
R^{(1)}_{\mu\nu} &=& \frac12 \left( \Box h_{\mu\nu} + \partial_\mu \partial_\nu h - \partial_\mu \partial_\alpha h^\alpha{}_\nu - \partial_\nu \partial_\alpha h^\alpha{}_\mu \right),
\\
R^{(1)} &=& \Box h - \partial_\mu \partial_\nu h^{\mu\nu},
\end{eqnarray}
where the wave operator is defined as $\Box := \eta^{\mu\nu} \partial_\mu \partial_\nu$ and the indices are raised or lowered using the Minkowski metric, such as $h^\alpha{}_\mu = \eta^{\alpha\beta} h_{\beta\mu}, \; h = h^\mu{}_\mu = \eta^{\mu\nu} h_{\mu\nu}$. Then the linear order of the Einstein equation,
\begin{equation}
G^{(1)}_{\mu\nu} = R^{(1)}_{\mu\nu} - \frac12 \eta_{\mu\nu} R^{(1)} = \kappa T^{(1)}_{\mu\nu},
\end{equation}
takes the form
\begin{equation} \label{eqBoxh}
\frac12 \left( -\Box \bar h_{\mu\nu}  - \eta_{\mu\nu}\partial_\alpha\partial_\beta \bar h^{\alpha\beta} +\partial_\mu \partial_\alpha \bar h^\alpha{}_\nu + \partial_\nu \partial_\alpha \bar h^\alpha{}_\mu \right) = \kappa T^{(1)}_{\mu\nu},
\end{equation}
where, as Einstein found in 1916, it is technically more convenient to reexpress the field equation in terms of the ``trace reversed'' tensor, $\bar h^{\alpha\beta}$, defined as\footnote{As he noted in 1918, the major mistake in his 1916 paper was the use of $\bar h_{\mu\nu}$ in place of $h_{\mu\nu}$ in the linearized material energy conservation expression.}
\begin{equation}
\bar h^{\alpha\beta} := h^{\alpha\beta} - \frac12 \eta^{\alpha\beta} h \quad \to \quad \bar h = \bar h^\alpha{}_\alpha = \eta_{\alpha\beta} \bar h^{\alpha\beta} = - h.
\end{equation}


As a generally covariant theory there is a gauge freedom associated with the freedom of coordinate transformations; this allows one to simplify the field equation with a suitable gauge choice. Consider the infinitesimal coordinate transformation $x'^{\mu} = x^\mu + \xi^\mu \; (|\xi^\mu| \ll 1)$, the metric is then transformed to
\begin{equation}
g'_{\mu\nu} = \frac{\partial x^\alpha}{\partial x'^\mu} \frac{\partial x^\beta}{\partial x'^\nu} g_{\alpha\beta} \quad \Rightarrow \quad g'_{\mu\nu} = g_{\mu\nu} - \partial_\mu \xi_\nu - \partial_\nu \xi_\mu.
\end{equation}
A convenient gauge choice for gravity is the harmonic gauge, $g^{\alpha\beta} \Gamma^\mu{}_{\alpha\beta} = 0$. The linearized version is analogous to the Lorenz gauge of electrodynamics:
\begin{equation}
\partial_\nu \bar h^{\mu\nu} = 0. 
\end{equation}
One can straightforwardly check that this is a valid attainable gauge. Consider the gauge condition written in terms of $h_{\mu\nu}$ subjected to a gauge transformation:
\begin{equation}
h'_{\mu\nu} = h_{\mu\nu} - \partial_\mu \xi_\nu - \partial_\nu \xi_\mu,
\end{equation}
then
\begin{eqnarray}
0 &=& \partial_\nu h'^{\mu\nu} - \frac12 \eta^{\mu\nu} \partial_\nu h'
\nonumber\\
&=& \partial_\nu ( h^{\mu\nu} - \partial^\mu \xi^\nu - \partial^\nu \xi^\mu ) - \frac12 \eta^{\mu\nu} \partial_\nu ( h - 2 \eta^{\alpha\beta} \partial_\alpha \xi_\beta)
\nonumber\\
&=& - \Box \xi^\mu + \partial_\nu  \bar h^{\mu\nu}.
\end{eqnarray}
This is a wave equation that can be solved for the gauge parameters $\xi^\mu$.

With this gauge condition the linear order field equation becomes a wave equation:
\begin{equation}
\Box \bar h_{\mu\nu} = 2 \kappa T^{(1)}_{\mu\nu}, \quad \Rightarrow \quad \Box h_{\mu\nu} = 2 \kappa \bar T^{(1)}_{\mu\nu}.
\end{equation}
The formal solution is given in terms of the retarded Green's function:
\begin{equation}
h_{\mu\nu}(\vec x, t) = \frac{\kappa}{2 \pi} \int d^3y \frac{\bar T^{(1)}_{\mu\nu}\left( \vec y, t - |\vec x - \vec y| \right)}{|\vec x - \vec y|}.
\end{equation}

In particular, let us consider the vacuum solutions  with $T_{\mu\nu} = 0$. A simple one is a plane gravitational wave:
\begin{equation}
h_{\mu\nu} = \epsilon_{\mu\nu} \mathrm{e}^{i k_\alpha x^\alpha},
\end{equation}
characterized by a 10 component constant polarization tensor $\epsilon_{\mu\nu} = \epsilon_{(\mu\nu)}$. The field equation gives the condition on the wave vector $k^\alpha = (\omega/c, \vec k)$:
\begin{equation}
\Box h_{\mu\nu} = 0 \quad \Rightarrow \quad k^\alpha k_\alpha = 0 \quad \Rightarrow \quad - (\omega/c)^2 + \vec k^2 = 0.
\end{equation}

Let us count the degrees of freedom of the gravitational wave. Indeed, the 4 conditions given by the Lorenz gauge
\begin{equation}
\partial^\mu \bar h_{\mu\nu} = 0 \quad \Rightarrow \quad k^\mu \epsilon_{\mu\nu} = \frac12 k_\nu \epsilon^\mu{}_\mu,
\end{equation}
do not completely fix all the  freedom. Actually it only fixes half of the freedom. So, there are 4 more additional conditions, and the true degrees of freedom are only 2,
\begin{equation}
10 \quad \longrightarrow \quad 6 \quad \longrightarrow \quad 2 \quad \textrm{(physical)}.
\end{equation}

The situation is analogous with that of electrodynamics. There one can choose the Lorenz gauge
\begin{equation}
\partial_\mu A^\mu = 0,
\end{equation}
for the electromagnetic wave in vacuum ($J_\mu = 0$). And the corresponding plane electromagnetic wave
\begin{equation}
A^\mu(x) = \epsilon^\mu \mathrm{e}^{i k_\alpha x^\alpha},
\end{equation}
is characterized by a 4-component polarization vector $\epsilon^\mu$. The Maxwell equation gives the null propagation condition
\begin{equation}
k_\alpha k^\alpha = - (\omega/c)^2 + \vec k \cdot \vec k = 0.
\end{equation}
It is well-known that the Lorenz gauge only fixes 1 among the 2 freedoms, namely
\begin{equation}
k^\mu \epsilon_\mu = 0,
\end{equation}
and the true degrees of freedom in an electromagnetic wave are 2,
\begin{equation}
4 \quad \longrightarrow \quad 3 \quad \longrightarrow \quad 2 \quad \textrm{(physical)}.
\end{equation}

\begin{table*}[ht]
\begin{center}
\begin{tabular}{|c||c|c|}
\hline
 & \mbox{\rule[0mm]{0cm}{.5cm} Electromagnetism} & Linearized Gravity \\
\hline\hline
 \mbox{\rule[0mm]{0cm}{.5cm} field} & $A_\mu$ & $h_{\mu\nu}$ (or $\bar h_{\mu\nu} $) \\
\hline
 \mbox{\rule[0mm]{0cm}{.5cm} source} & $J^\mu$ & $T^{(1)}_{\mu\nu}$ \\
\hline
 \mbox{\rule[0mm]{0cm}{.5cm} gauge transformation} & $A_\mu + \partial_\mu \lambda$ & $h_{\mu\nu} - \partial_\mu \xi_\nu - \partial_\nu \xi_\mu$ \\
\hline
 \mbox{\rule[0mm]{0cm}{.5cm} Lorenz gauge} & $\partial^\mu A_\mu = 0$ & $\partial^\mu \bar h_{\mu\nu} = 0$ \\
\hline
 \mbox{\rule[0mm]{0cm}{.5cm} field equation} & $\Box A_\mu = 4 \pi J_\mu$ & $\Box \bar h_{\mu\nu} = 2 \kappa T^{(1)}_{\mu\nu}$ \\
\hline
\end{tabular}
\end{center}
\end{table*}

Let us analyze the 4 additional freedoms in more detail. Consider the particular coordinate transformation with a 4-component parameter $\epsilon^\mu$ as
\begin{equation}
x^\mu \quad \to \quad x^\mu + i \epsilon^\mu \mathrm{e}^{i k_\alpha x^\alpha},
\end{equation}
then the polarization tensor is transformed as
\begin{equation}
\epsilon'_{\mu\nu} = \epsilon_{\mu\nu} + k_\mu \epsilon_\nu + k_\nu \epsilon_\mu, \qquad \epsilon'^\mu{}_\mu = \epsilon^\mu{}_\mu + 2 k^\mu \epsilon_\mu.
\end{equation}
One can straightforwardly check that this coordinate transformation indeed preserves the Lorenz gauge:
\begin{eqnarray}
k^\mu \epsilon'_{\mu\nu} &=& \underbrace{k^\mu \epsilon_{\mu\nu}}_{\frac12 k_\nu \epsilon^\mu{}_\mu} + \underbrace{k^2 \epsilon_\nu}_0 + k_\nu k^\mu \epsilon_\mu
\nonumber\\
&=& \frac12 (k_\nu \epsilon^\mu{}_\mu + 2 k_\nu k^\mu \epsilon_\mu) = \frac12 k_\nu \epsilon'^\mu{}_\mu,
\end{eqnarray}
and this remaining freedom allows us to impose additional conditions:
\begin{equation}
\epsilon^\mu{}_\mu = 0, \qquad \epsilon_{0 \mu} = \epsilon_{\mu 0} = 0.
\end{equation}
Therefore, there are only 2 independent degrees of freedom for gravitational radiation. All the gauge freedom is fixed  in the transverse-traceless (radiation) gauge,
expressed in terms of $h_{\mu\nu}$ by $\partial_\mu \bar h^{\mu\nu}=0$ and $h^\mu{}_\mu=0=h_{0\mu}$, consequently
\begin{equation}
h^{TT} = 0 = \bar h^{TT} \quad \Rightarrow \quad h_{\mu\nu}^{TT} = \bar h_{\mu\nu}^{TT}.
\end{equation}
In particular, a plane gravitational wave propagating in the $z$-direction, $\vec k = (0, 0, k) \to k^\mu = (k, 0, 0, k)$, can be expressed as
\begin{equation}
h^{TT}_{\mu\nu} = \left( \begin{array}{cccc} 0 & 0 & 0 & 0 \\ 0 & h_+ & h_\times & 0 \\ 0 & h_\times & -h_+ & 0 \\ 0 & 0 & 0 & 0 \end{array} \right) \mathrm{e}^{i k_\alpha x^\alpha},
\end{equation}
where the two polarizations are
\begin{equation}
\epsilon^{\mu\nu}_{(+)} = h_+ \left( \begin{array}{cccc} 0 & 0 & 0 & 0 \\ 0 & 1 & 0 & 0 \\ 0 & 0 & -1 & 0 \\ 0 & 0 & 0 & 0 \end{array} \right), \qquad
\epsilon^{\mu\nu}_{(\times)} = h_\times \left( \begin{array}{cccc} 0 & 0 & 0 & 0 \\ 0 & 0 & 1 & 0 \\ 0 & 1 & 0 & 0 \\ 0 & 0 & 0 & 0 \end{array} \right).
\end{equation}

\subsection{Gravitational Waves}
Let us discuss the two polarization modes of gravitational plane waves separately. Consider two nearby points separated by an infinitesimal coordinate displacement in the $x$-direction, i.e. $dx^\mu = (0, d\zeta, 0, 0)$; the corresponding proper separation length is
\begin{equation}
ds = \sqrt{g_{\mu\nu} dx^\mu dx^\nu} = \sqrt{g_{11}} \; dx^1 = \sqrt{\eta_{11} + h_{11}} d\zeta \sim \left( 1 + \frac{h_{11}}2 \right) d\zeta.
\end{equation}
Similarly, consider an infinitesimal coordinate separation in the $y$-direction, i.e., $dx^\mu = (0, 0, d\zeta, 0)$; the corresponding proper separation length is
\begin{equation}
ds = \sqrt{g_{\mu\nu} dx^\mu dx^\nu} = \sqrt{g_{22}} \; dx^2 = \sqrt{\eta_{22} + h_{22}} \; d\zeta \sim \left( 1 + \frac{h_{22}}2 \right) d\zeta.
\end{equation}
For the $+$-gravitational wave mode,
\begin{equation}
h_{11} = -h_{22} = h_+ \mathrm{e}^{i k_\alpha x^\alpha} \quad \to \quad h_+ \cos(\omega t - kz),
\end{equation}

\begin{figure}[ht]
\begin{center}
\includegraphics[height=3cm]{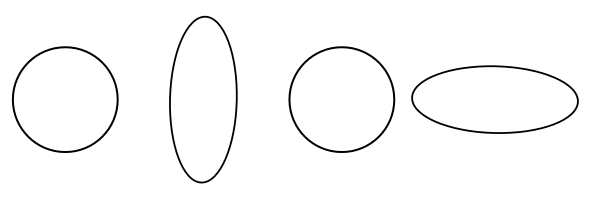}
\caption{The changing distances between a ring of test particles due to a GW $+$-mode propagating out of the page for, from left to right, $\omega t = 0, \pi/2, \pi, 3\pi/2$.} \label{FIG:GWplus}
\end{center}
\end{figure}

For the $\times$-gravitational wave mode, we should consider $dx^\mu = (0, d\zeta/\sqrt2, d\zeta/\sqrt2, 0)$ with the proper length
\begin{equation}
ds = \sqrt{g_{\mu\nu} dx^\mu dx^\nu} = \sqrt{\frac12 g_{11} + g_{12} + \frac12 g_{22}} \; d\zeta = \sqrt{\frac12 \eta_{11} + h_{12} + \frac12 \eta_{22}} \; d\zeta \sim \left( 1 + \frac{h_{12}}2 \right) d\zeta,
\end{equation}
and $dx^\mu = (0, d\zeta/\sqrt2, -d\zeta/\sqrt2, 0)$ with the proper length
\begin{equation}
ds = \sqrt{g_{\mu\nu} dx^\mu dx^\nu} = \sqrt{\frac12 g_{11} - g_{12} + \frac12 g_{22}} \; d\zeta = \sqrt{\frac12 \eta_{11} - h_{12} + \frac12 \eta_{22}} \; d\zeta \sim \left( 1 - \frac{h_{12}}2 \right) d\zeta.
\end{equation}
For the $\times$-mode we have
\begin{equation}
h_{12} = h_{21} = h_\times \mathrm{e}^{i k_\alpha x^\alpha} \quad \to \quad h_\times \cos(\omega t - kz),
\end{equation}

\begin{figure}[ht]
\begin{center}
\includegraphics[height=3cm]{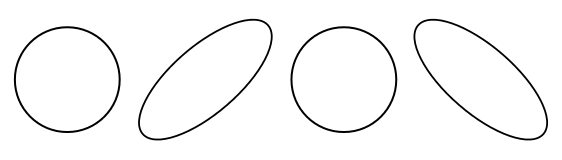}
\caption{The changing distance between a ring of test particles due to a GW $\times$-mode propagating out of the page for, from left to right, $\omega t = 0, \pi/2, \pi, 3\pi/2$.} \label{FIG:GWcross}
\end{center}
\end{figure}

Figures~\ref{FIG:GWplus} and~\ref{FIG:GWcross} show the changing distance between  a ring of test particles due to  GW $+$ and GW $\times$ modes propagating out of the page.

\subsection{Radiation from Localized Sources}
To discuss a gravitational wave generated by a source, we should consider the linear equation with source:
\begin{equation}
\Box \bar h_{\mu\nu} = 2 \kappa T^{(1)}_{\mu\nu}.
\end{equation}
The general solution formally can be expressed in terms of the retarded Green's function
\begin{equation}
\bar h_{\mu\nu}(\vec x) = \int d^3y G_\mathrm{ret}(\vec x - \vec y) 2 \kappa T^{(1)}_{\mu\nu}(\vec y) = \frac{\kappa}{2 \pi} \int d^3y \frac{T^{(1)}_{\mu\nu}(\vec y, t_\mathrm{ret})}{|\vec x - \vec y|},
\end{equation}
with the retarded time
\begin{equation}
t_\mathrm{ret} = t - \frac{|\vec x - \vec y|}{c}.
\end{equation}

Let us consider the solution for some special limits: (i) large distance: very far away from the source, i.e. $r \gg R$ where $R$ is the size of the source, such as a binary star system; (ii) slowly moving: radiation at low frequency $\lambda \gg \delta R \quad (\lambda = 2 \pi c/\omega)$, then (dropping here for simplicity the (1) label)
\begin{equation}
\bar h_{\mu\nu}(\vec x) \sim \frac{\kappa}{2 \pi |\vec x|} \int d^3y T_{\mu\nu}(\vec y, t_\mathrm{ret}).
\end{equation}

The conservation of source energy-momentum, i.e. $\partial_\mu T^{\mu\nu} = 0$ in the linearized theory, gives
\begin{eqnarray}
&& \partial_0 T^{00} + \partial_i T^{i0} = 0 \quad \to \quad \partial_0^2 T^{00} = - \partial_i \partial_0 T^{i0},
\\
&& \partial_0 T^{0i} + \partial_j T^{ji} = 0 \quad \to \quad \partial_0^2 T^{00} = \partial_i \partial_j T^{ji}.
\end{eqnarray}
For a localized source we have
\begin{eqnarray}
&& \int x^i x^j \underbrace{\partial_k \partial_l T^{kl}}_{\partial_0^2 T^{00}} d^3x = \underbrace{\int \partial_k (x^i x^j \partial_l T^{kl}) d^3x}_{\sim 0} - \int \left( x^j \partial_l T^{il} + x^i \partial_l T^{jl} \right) d^3x \sim 2 \int T^{ij}(x) d^3x
\nonumber \\
\Rightarrow && \int T^{ij} d^3x \sim \frac12 \partial_0^2 \int x^i x^j T^{00} d^3x \sim \frac1{2} \partial_t^2 \int \rho x^i x^j d^3x = \frac1{2} \frac{d^2}{dt^2} I^{ij}
\nonumber\\
\Rightarrow && \bar h^{ij} \sim \frac{\kappa}{2 \pi r} \left( \frac12 \frac{d^2}{dt^2} I^{ij} \right),
\end{eqnarray}
where the quadruple moment (2nd mass moment) is defined as $I^{ij} \equiv \int \rho x^i x^j d^3x$, with $\rho := T^{00}/c^2$. This is the well-known quadruple formula for a gravitational wave:
\begin{eqnarray}
&& \bar h^{ij} \sim \frac{\kappa}{2 \pi r} \left( \frac12 \frac{d^2}{dt^2} I^{ij} \right)
\nonumber\\
\Rightarrow && h^{TT}_{ij} \sim \frac{\kappa}{4 \pi r} \ddot Q^{TT}_{ij}(t - r/c), \qquad Q_{ij} := I_{ij} - \frac13 \delta_{ij} I^k{}_k.
\end{eqnarray}
(Here we just want to present the simple basic ideas; for careful calculations it is helpful to introduce a projection operator projecting from a general Lorenz gauge to  transverse-traceless modes for each plane wave~\cite{Maggiore}.)

Consider as an example a binary star system with the orbits for two stars $A$ and $B$ given by
\begin{eqnarray}
A: && x_a = R \cos\omega t, \quad y_a = R \sin\omega t,
\\
B: && x_b = - x_a, \quad y_b = - y_a.
\end{eqnarray}
The associated $tt$-component of the energy-stress tensor is
\begin{equation}
T^{00} = M c^2 \delta(z) \left[ \delta(x - R \cos\omega t) \delta(y - R \sin\omega t) + \delta(x + R \cos\omega t) \delta(y + R \sin\omega t) \right],
\end{equation}
and the components of the quadruple moment are
\begin{eqnarray}
&& I^{11} = 2 M R^2 \cos^2\omega t, \quad I^{12} = I^{21} = 2 M R^2 \cos\omega t \sin\omega t,
\nonumber\\
&& I^{22} = 2 M R^2 \sin^2\omega t, \quad I_{13} = I_{23} = I_{33} = 0.
\end{eqnarray}
This leads to the outgoing gravitational wave
\begin{equation}
\bar h^{ij} = \frac{\kappa M}{\pi r} \omega^2 R^2 \left( \begin{array}{ccc} -\cos2\omega t_\mathrm{ret} & -\sin2\omega t_\mathrm{ret} & 0 \\ -\sin2\omega t_\mathrm{ret} & \cos2\omega t_\mathrm{ret} & 0 \\ 0 & 0 & 0 \end{array} \right).
\end{equation}
One can compute the gravitational energy lost in the radiation. From the Einstein equation
\begin{equation}
G^{(1)}_{\mu\nu} = \kappa \left( T_{\mu\nu} + t_{\mu\nu} \right).
\end{equation}
To second order, one then gets an effective energy-momentum ``\textit{pseudo}tensor'' of the gravitational field:
\begin{eqnarray}
t_{\mu\nu} &=& \frac1{\kappa} \left( G^{(1)}_{\mu\nu} - G_{\mu\nu} \right) \sim {\cal O}(h^2)
\nonumber\\
&=& \frac1{\kappa} \left( \frac12 h_{\mu\nu} \eta^{\alpha\beta} R^{(1)}_{\alpha\beta} - \frac12 \eta_{\mu\nu} h^{\alpha\beta} R^{(1)}_{\alpha\beta} - R^{(2)}_{\mu\nu} + \frac12 \eta_{\mu\nu} \eta^{\alpha\beta} R^{(2)}_{\alpha\beta} \right).
\end{eqnarray}
In the transverse and traceless gauge it reduces to
\begin{eqnarray}
t_{\mu\nu} &=& \frac1{4 \kappa} \left( \partial_\mu h^{TT}_{\alpha\beta} \right) \left( \partial_\nu h_{TT}^{\alpha\beta} \right),
\\
h^{TT}_{ij} &=& \bar h^{TT}_{ij} = \frac{\kappa}{4 \pi r} \frac{d^2}{dt^2} Q^{TT}_{ij}(t - r/c),
\end{eqnarray}
and its average\footnote{This is sufficient for our needs here.  An effective gravitational energy-momentum tensor obtained by averaging for gravity waves was first proposed by Isaacson~\cite{Isaacson:1967zz,Isaacson:1968zza}, for detailed discussions see the textbooks~\cite{MTW,Maggiore}.}
is
\begin{equation} \label{taverage}
\langle t_{\mu\nu} \rangle = \frac1{4 \kappa} \langle (\partial h) (\partial h) \rangle = \frac1{4 \kappa} k_\mu k_\nu 2 (h_+^2 + h_\times^2) \langle \underbrace{\cos^2 k x \rangle}_{1/2}.
\end{equation}
According to the qudrupole formula, the power radiated by a binary system is
\begin{equation}
P = -\frac15 \frac{G_N}{c^5} \left\langle \frac{d^3 Q^{TT}_{ij}}{dt^3} \frac{d^3 Q^{TT}_{ij}}{dt^3} \right\rangle \simeq - \frac{128}{5} \frac{G_N}{c^5} M^2 R^4 \omega^6.
\end{equation}

\subsection{Sensitivity and Signal Strength}

Far from the GW sources, as it is in the present experimental/observational situations, the plane wave approximation is valid. Space averages can be replaced with time averages.
For our later discussion we will need some more technical details.
For orthogonal modes, the energy can be added in quadrature. For multi-frequency plane GWs, the total energy density in the spectral representation derived from~(\ref{taverage}), in terms of frequency $f = c k/(2 \pi)$, is~\cite{Kuroda:2015owv, Maggiore}
\begin{equation} \label{rhoc2a}
\rho c^2 = t_{00} = \frac{c^2}{16 \pi G_N} \int_{-\infty}^\infty (2 \pi)^2 f^2 \left( \left| {}^\mathrm{(f)}h_+(f) \right|^2 + \left| {}^\mathrm{(f)}h_\times(f) \right|^2 \right) df \equiv \int_0^\infty {}^\mathrm{(E)} S_h(f) df.
\end{equation}
${}^\mathrm{(E)}S_h(f)$ is defined as the (one-sided) energy spectral density of $h$ and is given by
\begin{equation} \label{defESh}
{}^\mathrm{(E)}S_h(f) = \frac{\pi c^2}{2 G_N} f^2 \left( \left| {}^\mathrm{(f)}h_+(f) \right|^2 + \left| {}^\mathrm{(f)}h_\times(f) \right|^2 \right) = \frac{\pi c^2}{8 G_N} f S_h(f),
\end{equation}
with
\begin{eqnarray}
S_h(f) &=& 4 f \left( \left| {}^\mathrm{(f)}h_+(f) \right|^2 + \left| {}^\mathrm{(f)}h_\times(f) \right|^2 \right) = f^{-1}(h_c(f))^2;
\nonumber\\
S_{hA}(f) &=& 4 f \left| {}^\mathrm{(f)}h_A(f) \right|^2 = f^{-1}(h_{cA}(f))^2 \quad \textrm{for a single polarization $A \, (A = +, \times)$},
\end{eqnarray}
are the spectral power density of $h$ and $h_A$, respectively, and
\begin{equation}
h_c(f) \equiv 2 f \left( \left| {}^\mathrm{(f)}h_+(f) \right|^2 + \left| {}^\mathrm{(f)}h_\times(f) \right|^2 \right)^{1/2}; \qquad h_{cA}(f) \equiv 2 f \left| {}^\mathrm{(f)}h_A(f) \right|,
\end{equation}
are the characteristic strains. For unpolarized GWs, $|{}^\mathrm{(f)}h_+(f)|^2 = |{}^\mathrm{(f)}h_\times(f)|^2$ and we have
\begin{equation} \label{ESh1}
{}^\mathrm{(E)}S_h(f) = \frac{\pi c^2}{G_N} f^2 \left| {}^\mathrm{(f)}h_+(f) \right|^2 = \frac{\pi c^2}{G_N} f^2 \left| {}^\mathrm{(f)}h_\times(f) \right|^2 = \frac{\pi c^2}{4 G_N} f S_{hA}(f).
\end{equation}
From~(\ref{defESh}), the energy density is proportional to $h_A^2$ for a particular polarization. General GWs can be resolved into a superposition of plane GWs, the formula (\ref{rhoc2a} -- \ref{ESh1}) are still applicable. For an early motivation and an in-step mathematical derivation, see, e.g.~\cite{Hellings:1983fr} and~\cite{Maggiore:1999vm}, respectively.

For background or foreground stochastic GWs, it is common to use the critical density $\rho_c$ for closing the universe as fiducial:
\begin{equation}
\rho_c = \frac{3 H_0^2}{8 \pi G_N} = 1.878 \times 10^{-29} \; \mathrm{g/cm^3},
\end{equation}
where $H_0$ is the Hubble constant at present. Throughout this article, we use the Planck 2015 value $67.8 (\pm 0.9) \, \mathrm{km \, s^{-1} \, Mpc^{-1}}$ for $H_0$~\cite{Ade:2015xua}. In the cosmological context, it is more convenient to define a {\it normalized GW spectral energy density} $\Omega_\mathrm{gw}(f)$ and express the GW spectral energy density in terms of {\it the energy density per logarithmic frequency interval divided by the cosmic closure density} $\rho_c$ for a cosmic GW sources or background, i.e.,
\begin{eqnarray}
\Omega_\mathrm{gw}(f) &=& \frac{f}{\rho_c} \frac{d\rho(f)}{df} = \frac{(\pi/8 G_N) f^3 S_h(f)}{3 H_0^2/8 \pi G_N} = (\pi^2/3 H_0^2) f^3 S_h(f)
\nonumber\\
&& \Bigl( = (2 \pi^2/3 H_0^2) f^3 S_{hA}(f) \quad \textrm{for unpolarized GW} \Bigr).
\end{eqnarray}
For the very-low-frequency band, the ultra-low-frequency band and the extremely-low-frequency band, this is a common choice.

\begin{table}[ht]
\begin{center}
\begin{tabular}{|c||c|c|c|}
\hline
 & Characteristic & Strain psd $[S_h(f)]^{1/2}$ & Normalized spectral \\
 & strain $h_c(f)$ &  & energy density $\Omega_\mathrm{gw}(f)$ \\
\hline\hline
 $h_c(f)$ & $h_c(f)$ & $f^{1/2} [S_h(f)]^{1/2}$ & $[(3 H_0^2/2 \pi^2 f^2) \Omega_\mathrm{gw}(f)]^{1/2}$ \\
\hline
 Strain psd $[S_h(f)]^{1/2}$ & $f^{-1/2} h_c(f)$ & $[S_h(f)]^{1/2}$ & $[(3 H_0^2/2 \pi^2 f^3) \Omega_\mathrm{gw}(f)]^{1/2}$ \\
\hline
 $\Omega_\mathrm{gw}(f)$ & $(2 \pi^2/3 H_0^2) f^2 h_c^2(f)$ & $(2 \pi^2/3 H_0^2) f^3 S_h(f)$ & $\Omega_\mathrm{gw}(f)$ \\
\hline
\end{tabular}
\end{center}
\caption{Conversion factors among the characteristic strain $h_c(f)$, the strain psd (power spectral density) $[S_h(f)]^{1/2}$ and the normalized spectral energy density $\Omega_\mathrm{gw}(f)$.} \label{TABLE:conversion}
\end{table}

\section{Do gravity waves exist?}
One of the most fascinating parts of Kennefick's book~\cite{Kennefick} is  Chapter 5.  Here we provide a brief summary of the highlights.

In 1936 Einstein submitted with Nathan Rosen a paper to the Physical Review entitled ``Do gravity waves exist''
which apparently came to a surprising conclusion: \emph{gravitational plane waves cannot exist}.

The Physical Review editor, John Tate, received a detailed referee report (for the complete referee report see Appendix A in~\cite{Kennefick}).
He sent it to Einstein with the request that he ``would be glad to have your reaction to the various comments and criticisms the referee has made''.

Einstein responded:

\begin{quotation}
Dear Sir,

We (Mr Rosen and I) had sent you our manuscript for \emph{publication} and had not authorized you to show it to specialists before it is printed.  I see no reason to address the---in any case erroneous---comments of your anonymous expert.  On the basis of this incident I prefer to publish the paper elsewhere.
\end{quotation}

Tate replied that he regretted Einstein's decision to withdraw the paper, but said he would not set aside the journal's review procedure.
Einstein's annoyance was such that he never published in Physical Review again.\footnote{Previously Einstein and Rosen had published in Physical Review two important works, one included the so-called ``Einstein-Rosen bridge'' (now referred to as a ``wormhole''), and the other the even more famous paper on quantum theory of Einstein, Podolsky and Rosen (EPR)---which was accepted by the editor without the advice of any referee.  But Tate's good instincts led him to seek in this case the opinion of a very suitable referee.}
In fairness to Einstein, such a review procedure was not then used by the German journals he had been using in the past.

Einstein then submitted the paper to the \emph{Journal of the Franklin Institute}, and it was accepted.
By this time Rosen had gone to Russia and Einstein had a new assistant, Leopold Infeld. Infeld received a visit from H. P. Robertson who had just returned from a year's sabbatical in Pasadena.  (Recently, i.e. only in this century, it came to light that Robertson was in fact the ``anonymous'' referee.) Robertson quickly befriended Infeld and soon got him to talking.  When told of the surprising new result of Einstein and Rosen,  Robertson expressed his skepticism. They went over the work together and Robertson soon persuaded Infeld of problems in the paper.  When Infeld went to tell Einstein, Einstein surprisingly said that he himself had just discovered some problems. The immediate consequence was that Einstein had to modify the talk he had prepared for the next day, at the end of which he said ``If you ask me whether there are gravitational waves or not, I must answer that I do not know.  But it is a highly interesting problem.'' Following this, the paper that had been accepted by the Journal of the Franklin Institute was revised by Einstein in proof to have rather the opposite of its original conclusion: it showed the existence of \emph{cylindrical} gravity waves.  It was published as ``On Gravitational Waves''~\cite{EinRos37}.

The abstract of the published paper reads:

\begin{quote}
The rigorous solution for cylindrical gravitational waves is given. For the
convenience of the reader the theory of gravitational waves and their production,
already known in principle, is given in the first part of this paper. After encountering
relationships which cast doubt on the existence of rigorous solutions for
undulatory gravitational fields, we investigate rigorously the case of cylindrical
gravitational waves. It turns out that rigorous solutions exist and that the
problem reduces to the usual cylindrical waves in euclidean space.
\end{quote}

And it ends with a note:

\begin{quote}
Note.--The second part of this paper was considerably altered by me after
the departure of Mr.~Rosen for Russia since we had originally interpreted our
formula results erroneously. I wish to thank my colleague Professor Robertson
for his friendly assistance in the clarification of the original error. I thank also
Mr. Hoffmann for kind assistance in translation.
\end{quote}

Einstein and Rosen were looking for an exact plane wave solution, but actually they found cylindrical waves, they are often referred to as Einstein-Rosen waves.\footnote{Cylindrical waves had been found earlier by Beck in 1925~\cite{Beck25}, but it seems nobody knew of this work.  Regarding plane waves, Baldwin \& Jeffery in 1926~\cite{BaldJeff26} had already established that one could not give a single coordinate system singularity free description, but Einstein did not pay much attention to the literature. A definitive analysis of plane waves was done much later~\cite{Bondi57, BondiPR59}.}

What Einstein and Rosen discovered was that it was not possible to construct a metric representing plane gravitational waves in a given coordinate system which did not include a singularity somewhere. Much later it was shown that this singularity is merely a coordinate singularity, not a physical singularity, but but in those days there was no well developed technique to distinguish the two and Einstein and Rosen had interpreted it as physical and, consequently, argued that such spacetimes could not exist. The published version also discusses the possibility that binary stars would not emit gravitational waves, even though the quadrupole formula suggests that they would (this foreshadows future controversies still to come).

Rosen did not know of the revision or publication until a friend had sent him a newspaper clipping, meanwhile he had also found some error, but was not happy with the revisions made by Einstein. His skepticism long continued. He published a paper in a Soviet journal in 1937 arguing that gravitational waves do not exist because of a singularity in the solution~\cite{Rosen37}. In 1955 at a conference he argued that gravitational waves cannot transport energy~\cite{Rosen56}. And in 1979 he published a paper ``Does Gravitational Radiation Exist?''~\cite{Rosen79}.

\section{Some field theory considerations }
Ever since Faraday and Maxwell we think in terms of interaction \emph{fields}, locally the movement of a source affects an interaction field, the effects on the  field propagate to distant locations where they can produce local effects.  Thus for a field theory describing a fundamental interaction one naturally expects propagating waves.

Maxwell's electrodynamics is now our prototype for physical interactions. To understand some other phenomenon, including gravity, we look to electromagnetism for an analogous situation.
For example, one can consider the simple argument due to J.~J.~Thompson (see Fig.~4.6 in Ref.~\cite{MTW}) of the outwardly propagating changes in a Coulomb field when a charge is moved from one point to a nearby point. The argument vividly accounts for the $1/r$ falloff of the generated radiation as well as its angular dependence.  If one similarly contemplates suddenly changing the distribution of some masses in a way that changes the quadrupole moment---just from considering the Newtonian field---one can likewise clearly see that there should analogously be an outgoing pulse of changing field, of a quadrupole nature, with $1/r$ falloff---``gravitational radiation''.  If the mass distribution has a periodically changing quadrupole moment then surely one would have outgoing ``gravitational waves''.  How similar are these non-linear ``waves'' of GR to those of linear electrodynamics? As we shall see, there have been many controversies regarding derivations of the Einstein quadrupole radiation formula (for a very brief derivation see Landau \& Lifshitz~\cite{LL}), the differences largely depended on the extent to which people trusted the electromagnetic analogy.

\section{Two conferences: Berne 1955 \& Chapel Hill 1957}
The period from 1925 to 1955 has been described as ``The low water mark of general relativity''~\cite{Eisenstaedt}.  Kennefick associates the renaissance of GR  with two conferences, which are credited with initiating the regular series of international GR conferences.  The first was at Berne in 1955~\cite{Berne55}.  One of the major new ideas was promoted there by Felix Pirani (apparently inspired by J.~L.~Synge), namely the formula for \emph{geodesic deviation} (often referred to in terms of its Newtonian analogue, the \emph{tidal force}). This is the key relation that governs the response of gravitational wave detectors; it is the analogue of the Lorentz force in electromagnetism:
\begin{equation}
\frac{D v^\mu}{D \tau} = \frac{q}{m} F^\mu{}_\nu v^\nu.
\end{equation}
For two nearby ``test'' particles moving on geodesics,
\begin{equation}
0=\frac{D v^\mu}{D \tau} := \frac{d v^\mu}{d \tau} + \Gamma^\mu{}_{\alpha\beta} v^\alpha v^\beta, \qquad v^\mu = \frac{d x^\mu}{d \tau},
\end{equation}
the vector describing their relative separation has the acceleration
\begin{equation}
\frac{D^2 \xi^\mu}{D \tau^2} = R^\mu{}_{\alpha\beta\nu} v^\alpha v^\beta \xi^\nu. \label{deviation}
\end{equation}
Aside from the number of components in the curvature vs the Maxwell tensor (and the lack of a factor like the electromagnetic charge-to-mass ration), the main difference is that in the electromagnetic case the field can cause charges to accelerate away from inertial motion, whereas in the gravitational case two inertially moving particles have a relative ``tidal'' acceleration.
By measuring the relative displacement of nearby particles one can find all the components of the curvature tensor.  Indeed one could take the \emph{geodesic deviation equation}~(\ref{deviation}) as a definition of curvature.  Rather like the electromagnetic Lorentz force law, it can serve as both a ``definer of fields'' and a ``predicter of motions'' (see Box.~3.1 in Ref.~\cite{MTW}). The geodesic deviation equation gives an alternative way to understand the relative motion of a ring of test particles acted on by a GW, as depicted in Figs.~\ref{FIG:GWplus} and~\ref{FIG:GWcross}.

Another noteworthy incident happened at this Berne conference.
It was at this conference in 1955 that Marcus Fierz said (prophetically) to Bondi ``The problem of gravitational waves in general relativity is now ripe for solution and you are the person to solve it.''~\cite{Bondi87}.

Even more influential was the conference in 1957 at Chapel Hill, North Carolina, organized by Cecile DeWitt-Morette and Bryce DeWitt~\cite{ChapelHill}. Later, several participants stated that their most vivid recollection was of Herman Bondi, actively illustrating  how his changing quadrupole moment would generate gravitational waves when he exercised waving his two dumbbells. Another notable highlight was the so-called ``sticky bead argument''.  Apparently it was independently due to Bondi and Richard Feynman (who registered under the pseudonym ``Mr. Smith'', and had arrived a day late, missing Bondi's presentation).  The idea is that a passing gravitational wave, producing geodesic deviation, would cause beads sliding on a rod with some friction to heat the rod.  The point is that the energy for heating the rod must come from the gravitational wave.  Many have accepted this as a convincing argument that \emph{gravitational waves can transport energy}.  But unlike electromagnetism, they are not dipole but rather quadrupole waves.

Furthermore from about that period of time, various astrophysical discoveries and new theoretical ideas, like quasars, pulsars, gravitational collapse and black holes, encouraged an increasing interest in emissions from strongly gravitating systems.

\section{Skeptics and non-skeptics}
Kennefick has an interesting classification of the people working on gravitational wave theory into two major groups, labeled \emph{skeptics} and {\emph{non-skeptics}.
The skeptic label is not just for those (hardly anyone) who really doubted that gravitational waves exist.  Rather the key point concerns the electromagnetic analogy.  He uses the skeptic label for those who did not trust the electromagnetic analogy, who would use it but would be looking for significant differences,  places where it would break down.  For him non-skeptics are those who regarded the analogy as quite reliable.  Both camps might even do exactly the same calculation: when considering a gravitational problem they both might try the approach that worked for an analogous situation in electromagnetic theory.  But one would be looking for a difference while the other expected that it would work very much the same way.  Both attitudes could be highly productive.  He notes in particular Wheeler and Bondi,  two who probably contributed more to our understanding than any others, Wheeler he classifies as a non-skeptic and Bondi as a skeptic. Kennefick's list of non-skeptics include: Lev Landau, John Wheeler, Felix Pirani, and Andrzej Trautman.
His list of skeptics includes A.~S.~Eddington, Nathan Rosen, Leopold Infeld, Peter Havas, Herman Bondi, and, more recently, Fred Cooperstock~\cite{Cooperstock}.

Along with their different attitudes towards the electromagnetic analogy, many of the skeptics had doubts as to whether gravitational waves could transport energy, and, especially, whether gravitational binaries emitted gravitational waves that transported energy.


Here from Kennefick's book is an example of the types of arguments used:  ``When interest in general relativity began to pick up again in the mid-fifties, Rosen and Infeld advanced a number of arguments whose common point was that binary star systems would not undergo orbital decay as a result of emitting gravitational waves. Hermann Bondi also entertained serious doubts on this score, arguing that the analogy with electromagnetism which lay behind the original notion of gravitational waves, actually pointed this way. His view was that in electrodynamics it was believed that accelerating charges emitted radiation and that the same was expected to hold true in the case of gravity. But since the theory was a theory of general relativity, how did one define what was accelerating? In Bondi's view, an inertial particle in general relativity was one which followed a geodesic. An accelerating particle was one which did not. Since binary stars in orbit around each other followed the geodesics of the local spacetime, they were not accelerating, in this sense. As particles in a form of inertial motion, their motion would not be of the type which should decay in response to radiation reaction.''
Also ``Bondi thought that two orbiting bodies consisting of pressure-free dust would not radiate, since every particle contained in the two bodies would follow a geodesic throughout their motion, \dots''~\cite{Kennefick} page 200.

We can add some new people to the list of skeptics.  Aldrovandi, Pereira \& Vu have argued that non-linear effects may be more important than had been realized, perhaps affecting the observable polarizations~\cite{Aldrovandi:2007cd, Aldrovandi:2008ci}.
And in a work just published Zhao-Yan Wu writes: ``\dots we show that gravitational field does not exchange energy-momentum with particles. And it does not exchange energy-momentum with matter fields either. Therefore, the gravitational field does not carry energy-momentum, it is not a force field and gravity is not a natural force.'' and ``The gravitational field does not exchange energy-momentum with both particles and electromagnetic field. So, it does not carry energy-momentum. It is not a force
field. And gravity, the oldest natural force known to people, is not really a natural force.''  also ``GR has been the most beautiful theory in physics, but it was messed by pseudotensors, non-localizability and gravitational energy-momentum which resides nowhere like a ghost.''~\cite{Wu2016}

\section{Do gravitational waves carry energy?}
Whether gravitational waves carry energy has been debated since the beginning. The issue of gravitational energy is not so simple. From the work of Hilbert, Klein and Noether it was established that there is no proper energy density for general relativity (or indeed for any generally covariant theory of gravity); for an extensive discussion of this point see Ref.~\cite{GR100}.  For the density of gravitational energy-momentum Einstein had found a pseudotensor expression, i.e., not a proper tensor but rather an expression that was inherently reference frame dependent.  Einstein had used his pseudotensor and found that the transverse-transverse waves carry energy. But many objected to using this pseudotensor to describe gravitational energy, including Levi-Civita, Bauer~\cite{Bauer1918}, and Schr\"odinger~\cite{Schroedinger1918} (years later in his book~\cite{Schroedinger1950} he calls it ``sham'').  Eddington and Pauli both argued that gravitational energy was non-localizable. Nowadays gravitational energy is regarded as being \emph{quasi-local}---associated with a closed 2-surface (see~\cite{Sza09} for a comprehensive review of quasi-local energy-momentum.)

Regarding gravitational energy, there are two main ambiguities: (1) there is no unique expression, (2) each expression depends on the choice of reference frame and there is no preferred reference frame  (both of these issues are addressed in our Hamiltonian boundary approach~\cite{GR100}).  Because of this feature, different calculations making different reasonable choices can give different results.  To appreciate the meaning of such results one must keep in mind that there really are various legitimate ways to measure energy, not one unique formula.  (Actually this is case in physics in general, for example in thermodynamics the internal energy, the enthalpy, and the Helmholtz and Gibbs free energy are all physically meaningful.)  Not long ago the energy of an exact gravitational plane wave was calculated using the teleparallel equivalent of GR (one of the main reasons for interest in this alternative approach to GR is precisely because the teleparallel formulation is supposed to have advantages for describing gravitational energy), using an obviously reasonable reference frame~\cite{Obukhov:2009gv}.  It was found that, according to this measure, plane waves do not transport energy. The authors say ``Our results do not mean at all that gravitational waves cannot exist. However, they seem to support the idea that the true gravitational wave should be essentially a nonlinear physical phenomenon. The analysis of the specific example of the exact plane wave considered here shows that plane waves seem to have rather unusual properties, such as the vanishing of their energy-momentum current.''
We have no quarrel with this calculation, but expect that other alternative, no less legitimate, energy expressions could well give a different answer.

In the early days some (including Klein, Levi-Civita and Lorentz) argued that the \emph{proper gravitational energy density} is just the Einstein tensor. Then total energy is rather trivially conserved, it vanishes everywhere:
\begin{equation}
-\frac{1}{\kappa} G_{\mu\nu} + T_{\mu\nu} = 0.
\end{equation}
In particular, in the vacuum, although curvature waves propagate, their proper energy-momentum density vanishes. According to this idea gravitational waves waves do not carry energy.  This would have important consequences: such waves could not be detected by resonant mass detectors. But detectors based on interferometry could work, since they do not depend on absorbing energy, they directly detect changes in length. In our understanding from Hilbert, Noether and our Hamiltonian approach this idea is in fact quite correct---but a density is not the whole story. There is more to energy-momentum---associated with any region there is also a (generally non-vanishing) surface term: gravitational energy is \emph{quasi-local}~\cite{GR100}.

A more recent proposal has some features that are similar to the aforementioned early idea. In 1992 Cooperstock~\cite{Cooperstock} proposed his \emph{new hypothesis} according to which the energy-momentum conservation laws are devoid of content in vacuum, energy is localized where the energy-momentum tensor is nonvanishing, consequently gravitational waves are not carriers of energy in vacuum.  This led to later work with Dupre in which they advocated an invariant spacetime energy expression based on the Ricci tensor and the Tolman integral~\cite{Cooperstock:2010zz, Cooperstock:2013joa}.

Below we will present some weighty arguments that support the idea that gravitational waves can transport energy.  However, although there are many conceptions that may be satisfactory for certain practical purposes, it should be kept in mind that even after a century there is no consensus on any specific gravitational energy expression.

Bondi has given a strong argument for \emph{inductive energy transport} through empty space via the gravitational field in both Newtonian theory~\cite{BondiMcC60} and in GR~\cite{Bondi}.
There has long been observational evidence for such an energy transfer.  One example is the slowing of the Earth's rotation due to tidal effects.  A cleaner and more dramatic example is offered by the volcanoes of Jupiter's moon, Io.  One might have expected Io to rather cold and geologically dead, like our moon.  What provides the necessary heat for the volcanoes?  The answer is tidal heating due to the effect of Jupiter.  The computation has been done using Newtonian ideas~\cite{Purdue} and in GR via both pseudotensors~\cite{Purdue,Favata} and quasi-local energy approaches~\cite{BoothCr00}.

Inductive energy transfer seems solidly established.  However whether energy can actually be transported via gravitational waves is another issue.
Bondi produced his celebrated \emph{news function}~\cite{Bondi_vdBurgMetz62}\footnote{Bondi has described this paper as ``the best scientific work I have ever done''\cite{Bondi90}.}  to provide a clear indication.  Essentially it measures the shear of the outgoing wavefront.  As mentioned earlier, a widely used measure of the energy-momentum in a gravitational wave is due to Isaacson~\cite{Isaacson:1968zza}.  He produced a proper tensorial quantity obtained by averaging over several wavelengths that describes the energy in an invariant way.

Christodoulou in 2009~\cite{Christodoulou09} has provided perhaps the most compelling result.  He proved that asymptotically flat vacuum initial data could be such that it would in the future form a trapped surface. Essentially this means that gravitational waves could be focused to form a black hole. So certainly gravitational waves carry energy, since they can condense and act like a ``mass'' (Wheeler had asked him to try to show this about 40 years earlier.)

\section{The problem of motion and the quadrupole formula}
Kennefick remarks ``A great deal of confusion arose out of attempts to calculate, via the problem of motion, the rate of energy loss in binary star systems, due to gravitational wave emission, as Eddington had done for the case of a rotating rod. \dots Between 1947 and 1970, dozens of different calculations by a wide variety of methods gave an almost equally wide variety of results.  Not only did many calculations not recover the quadrupole formula at the leading order, but some (principally those due to Infeld and collaborators\footnote{See Refs.~\cite{InfeldS, Rosen56, InfeldP}}) found no emission, and three separate publications (Hu 1947; Peres 1959; Smith and Havas 1965)\footnote{See Refs.~\cite{Hu, Peres, SmithHavas}.} found that the binary would paradoxically gain energy as the result of emitting gravitational waves.''~\cite{Kennefick} p~125.

Regarding the approach to the problem of motion taken by Einstein, Infeld and Hoffman (EIH)~\cite{EIH}, Kennefick (\cite{Kennefick} page 146) notes that
\begin{quotation}
``James Anderson (1995) has insisted that most subsequent attempts to extend the problem of motion, especially in the direction of radiation reaction, \dots
, failed to appreciate or take advantage of the best points of the EIH scheme.  In his view, EIH, which he counts among Einstein's most significant work, was the great lost scheme of the postwar period, \dots .''
``The extent to which EIH has received a very mixed press is indicated by the marked disagreement between Havas and Anderson, the former decrying its malign influence on the field, the latter lamenting its lack of influence.  Havas was and is a trenchant critic of EIH, whereas Anderson now regards it as the most significant work on the problem of motion in general relativity.'' (page 147)
\end{quotation}

Anderson viewed the EIH papers as ``arguably one of Einstein's greatest contributions to physics'',
``these papers contain what I would claim is one of Einstein¡¦s greatest
contributions to physics and the only reliable method to date for deriving conditions of motion.''~\cite{Anderson}

Regarding the quadrupole formula, the skeptics can make a good case.  The electromagnetic analogy may not, after all, be such a good guide for the radiation reaction of gravitating bodies.  Accelerated charges radiate, but there are well-known problems with obtaining a suitable expression for the associated electromagnetic radiation reaction force.  For point charges the ``standard'' expression, due to Abraham-Lorentz-Dirac has various pathologies.\footnote{With a 3rd order equation of motion, causality (pre-acceleration), and/or runaway solutions~\cite{Roh65}.} Indeed Anderson has turned the electromagnetic analogy around and addressed the electromagnetic radiation reacting  problem by taking a GR inspired EIH approach~\cite{Anderson}.

The quadrupole formula for gravitating binaries was for a long time quite controversial.
It involved quite complicated calculations. Kennefick describes J\"urgen Ehlers as an agnostic who pushed people to work on  clarifying the issues. Over the years many people made contributions to clarifying things.  In addition to those people and works discussed, we should also mention Bonner, Fock, Robinson \& Trautman~\cite{RobTraut60}, Plebansky, and Burke \& Thorne~\cite{BurkeThorne}.

Significant issues include: (i) using advanced vs retarded vs half advanced--half retarded propagators, (ii) boundary conditions.  Boundary conditions are needed for both the inner strong field region and the asymptotic wave zone. It took a long time to develop the necessary technique of matched asymptotic expansions. Until then many different answers were obtained---some, as we noted, even gained energy. There were two main approaches: the post-Newtonian, called \emph{slow} because it was based on an expansion in powers of a small $v/c$, and the so-called \emph{fast} approach, based on the linearized theory, with no assumption of a small $v/c$.

Kennefick's Figure 10.7 is a table labeled ``The quadrupole formula controversy''. Here we present some of the entries along with the appropriate references to the literature.

\begin{small}
\begin{itemize}
\item[1976] Ehlers, Rosenblum, Goldberg and Havas dispute the validity of the various derivations of the quadrupole formula for binary stars.~\cite{EhlersRGH76}
\item[1978] Taylor and collaborators announce that the orbital decay of PSR 1913+16 is observed to be in agreement with the prediction of the quadrupole formula.~\cite{Taylor80}.
\item[1980]  Martin Walker and Clifford Will propose their three iterations test of the validity of quadrupole formula derivations.~\cite{WalkerWill}
\item[1981] Rosenblum publishes fast-motion scattering calculation disagreeing with the quadrupole formula.~\cite{Rosenblum}
\item[1982] Cooperstock and Hobill argue against Walker and Will's thesis that the history of the controversy is essentially settled.~\cite{CooperstockHobill}
\item[1983] Darmour's verdict on the binary pulsar data's agreement with theory and the derivation of the quadrupole formula.~\cite{Darmour}
\end{itemize}

\end{small}

By 1985 the controversy was effectively over. To provide just a small taste of how things went, we quote a few passages from Kennefick~\cite{Kennefick}:


\begin{quotation}
``Darmour's detailed analysis of the problem of motion, intended to compare directly with the experimental results from PSR 1913+16, is the closest thing to a solution to the quadrupole motion dispute, in the sense of its wholly or partially satisfying as many people as possible. Anderson's approach is regarded by a number of authorities as the most accessible and direct derivation of the quadrupole formula, \dots'' (page 248)
\end{quotation}

\begin{quotation}``[Anderson] was quite critical of other procedures that purported to derive the quadrupole formula.  He and Darmour were highly crital of each other's calculations, \dots Each was inclined to regard his own contribution as the only correct derivation of the equations of motion for radiating systems.'' (page 251)
\end{quotation}

\begin{quotation}
``The radiation reaction work in the 1970s and '80s differed in one essential respect from that of previous decades.  In most cases, the results of the various papers published agreed with each other and with the quadrupole formula.'' (page 252)
\end{quotation}

\section{The binary pulsar}
The binary pulsar PSR B1913+16 was discovered in 1974. In 1993 Taylor and Hulse were awarded the Nobel Prize for their work. In 1978 Taylor first announced that the orbital parameters of the binary pulsar were decaying in accordance with the quadrupole formula~\cite{Taylor:1979zz, Taylor80}. More recent data shows that the decay fits the quadrupole formula very well for over three decades~\cite{Weisberg:2010zz}. Fig.~\ref{FIG:BPulsar} shows a comparison of the observed and predicted orbital period decay in PSR B1913+16.

\begin{figure}[ht]
\begin{center}
\includegraphics[width=3in]{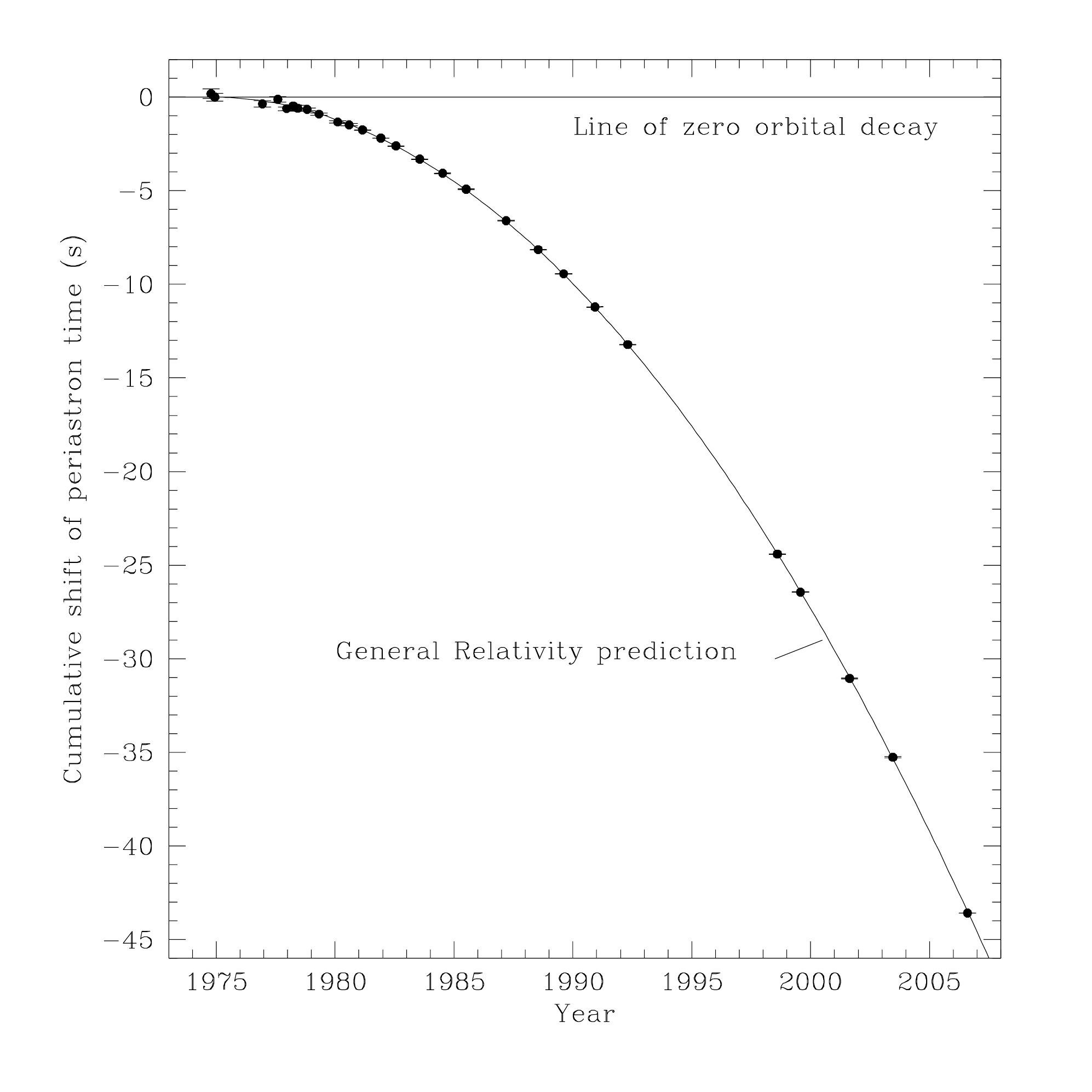}
\caption{Comparison of the observed and predicted orbital period decay in PSR B1913+16. The orbital decay is quantified by the shift in the time of periastron passage with respect to a nondecaying orbit. The parabolic curve is the predicted decay from GR~\cite{Weisberg:2010zz}.} \label{FIG:BPulsar}
\end{center}
\end{figure}

This is indeed good, although indirect, evidence for the existence of gravitational waves.  Furthermore it is good evidence that (i) gravitational waves do carry energy, and  (ii) the quadrupole formula is valid for gravitating binaries.

Based on more than thirty-two years (from 1974 through 2006) of timing observations of the relativistic binary pulsar B1913+16, the cumulative shift of peri-astron time is over 43 s. The calculated orbital decay rate in general relativity using parameters determined from pulsar timing observations agreed with the observed decay rates. From this and a relative acceleration correction due to solar system and pulsar system motion, Weisberg, Nice and Taylor~\cite{Weisberg:2010zz} concluded that the measured orbital decay to the GR predicted value from the emission of gravitational radiation is $0.997 \pm 0.002$ providing conclusive evidence for the existence of gravitational radiation. 

Evidence has also come from observation of other binary pulsar systems. Kramer et al.~\cite{Kramer:2006nb} did an orbit analysis of the double pulsar system PSR J0737-3039A/B from 2.5 years of pulse timing observations and found that the orbit period shortening rate 1.252(17) agreed with the GR prediction of 1.24787(13)
to within a ratio of 1.003(14).
Freire et al.~\cite{Freire:2012mg} analyzed about 10 years of timing data of the binary pulsar J1738+0333 and obtained the intrinsic orbital decay rate to be $(-25.9 \pm 3.2) \times 10^{-15}$, which agrees well with the calculated GR value $(-27.7^{+1.5}_{-1.9}) \times 10^{-15}$ using the determined orbital parameters. Further precision and many more systems are expected in the future for observable GW radiation reaction imprints on orbital motion.

\section{Weber Bar Experiment}
As we mentioned in the INTRODUCTION, the detection gap between astrophysical GW source strength and the technological achievable sensitivity was about 15 -- 16 orders of magnitude in amplitude in the low-frequency band and high-frequency GW band 100 years ago. The experimental efforts to detect gravitational waves did not begin until Joseph Weber, his post-doc David Zipoy and student Robert Forward started to work on GW experiments from 1958 on. They were pushing the technology for doing such an experiment at that time.\footnote{Weber wrote the first textbook on gravitational waves, it was published in (1961)~\cite{Weber61}.}

In 1966, fifty years after Einstein's first paper on GWs and fifty years before the LIGO announced their first detection of GWs, Weber published a paper~\cite{Weber:1966zz} on the sensitivity of the detector ``Observation of the Thermal Fluctuations of a Gravitational-Wave detector''. In the paper, he stated that strains as small as a few parts in $10^{16}$ are observable for a compressional mode of a large cylinder. This is an important milestone and narrows the 16 orders of gap of GW strength and experimental sensitivity to a 6 orders gap in the GW detection. Four months later, Weber published another paper ``Gravitational radiation''~\cite{Weber:1967jye} and reported that: ``The results of two years of operation of a 1660-cps gravitational-wave detector are reviewed. The possibility that some gravitational signals may have been observed cannot completely be ruled out. New gravimeter-noise data enable us to place low limits on gravitational radiation in the vicinity of the earth's normal modes near one cycle per hour, implying an energy-density limit over a given detection mode smaller than that needed to provide a closed universe.'' Setting upper limits on GWs has become a standard practice in reporting most GW experimental results up to the recent discovery.

In order to calibrate their experiment, Sinsky and Weber (Sinsky 1968~\cite{Sinsky:1968}; Sinsky and Weber 1967~\cite{Sinsky:1967}) used the dynamic gravitational field of an acoustic oscillating bar with an oscillation frequency 1660 Hz and dynamic strains around $10^{-4}$ to induce a resonant oscillation of the bar GW detector of strain amplitude of about $10^{-16}$ (Figure~\ref{FIG:detector} from Physics Today~\cite{Weber:1968a}). The generator bar and the detector bar are in separate vacuum tanks. Experimental results are in agreement with theoretical calculation to within 3\%.

As we have seen in the introduction this achievement in strain sensitivity is an important milestone for GW detection since it bridged the 16-order gap between source strength on earth and detector sensitivity to 6 orders only.

\begin{figure}[ht]
\begin{center}
\includegraphics[width=4in]{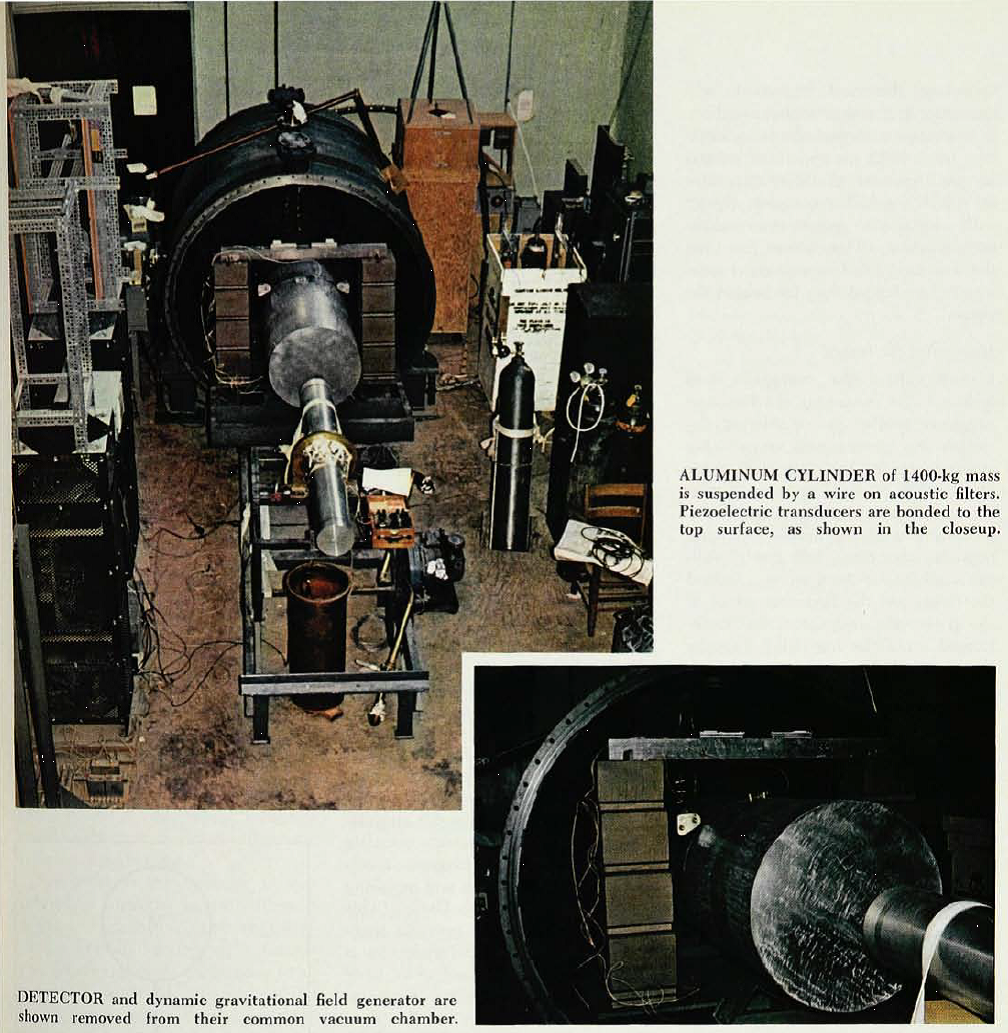}
\caption{Sinsky's calibration (from Physics Today~\cite{Weber:1968a}).} \label{FIG:detector}
\end{center}
\end{figure}

The 1968 discovery of pulsars gave hope of the existence of stronger GW sources. Binary pulsars (neutron stars) and their merging have served as light houses (fiducial standards for GW strains) for ground GW detectors.

In 1969, Weber announced evidence for the detection of gravitational radiation and believed that he had observed coincidences on gravitational-radiation detectors over a base line of about 1000 km at Argonne National Laboratory and at the University of Maryland. Weber listed 17 coincident events in this announcement~\cite{Weber:1969bz}.

In 1970, Weber~\cite{Weber:1970mx} reported observing a large anisotropy for gravitational-radiation-intensity as a function of sidereal times with peaks in the direction of the galactic center and in the opposite direction consistent with 12-h antenna symmetry. He concluded that ``The large (exceeding 6 standard deviations) sidereal anisotropy is evidence that the gravitational-radiation-detector coincidences are due to a source or sources outside the solar system. The location of the peaks suggests that the sources is the $10^{10}$ solar masses at the galactic center.''

Weber's results aroused great interests in GW detection. However, the theoretical studies using general relativity found that it is difficult to reconcile with Weber's finding at that time. (See, Press-Thorne 1972 for a review~\cite{Press:1972am}.)

Weber's results stimulated other experimental groups to start their own experiments. Their first results follow:

(i) In Moscow State University (MSU), Braginskii, Manukin, Popov, Rudenko and Khorev~\cite{Braginskii:1972db} constructed two GW antennas (one in MSU, the other 20 km apart at the Space Research Institute [SRI] with the same parameters as those of Weber~\cite{Weber:1969bz, Weber:1970mx}. Instead of using piezoelectric pickups, they used modulated capacitive displacement pickups. The capacitive pickup transformed a vibration amplitude of  $\sim 4.5 \times 10^{-16} \, \mathrm{m}$ (corresponding to the rms amplitude $\sigma_\mathrm{Brown}$ of the Brownian oscillations) into a radio-frequency signal of amplitude $\sim 4 \times 10^{-7} \, \mathrm{V}$. (For comparison, the Brownian fluctuations corresponded to a piezoelectric-pickup signal level of $5 \times 10^{-10} \, \mathrm{V}$ in the Weber (1969, 1970) experiment.) The construction and calibration of the pickup is described in Braginskii, Mitrofanov, Rudenko and Khorev~\cite{Braginskii:1971}. With their experiment, Braginskii et al. concluded ``The decrease in the number of coincidences following the introduction of delay in one of the channels, and the anisotropy of the distribution of the coincidences in sidereal time, are significant arguments in favor of the correlated bursts in Weber's experiments. The absence of coinciding bursts above the $3 \sigma_\mathrm{Brown}$ level in our experimental scheme does not contradict the astrophysical estimates~\cite{Braginskii:1969}.''

(ii) In Bell Laboratories, Tyson~\cite{Tyson:1973ra} built a calibrated detector of kilohertz-band gravitational radiation which had improved sensitivity over Weber's apparatus. No events were observed at 710 Hz during 3 month's observation.

(iii) Levine and Garwin (Levine and Garwin 1973~\cite{Levine:1973sr}; Garwin and Levine 1973~\cite{Garwin:1973wk}) built a bar GW detector at 1695 Hz and did not find any gravity-wave signals.

(iv) Data from a bar gravitational-wave detector constructed by Drever's group in Glasgow University suggest that pulses of gravitational radiation of less than a few milliseconds duration are too infrequent to account for signals reported by Weber in 1970~\cite{Drever:1973xf}.

In summary, the results of these four experiments did not confirm Weber's observation of gravitational waves. Further experimental results and better constraints on the intensity of gravitational radiation based on the coincidence of detection of a pair of bar antennas came in~\cite{Braginskii:1974, Billing:1975, Douglass:1975qs, Hirakawa:1975qf}. By 1975, almost all people in the GW detection community agreed that Weber had not detected gravitational waves. Nevertheless, the search for gravitational waves continued with even more intensity.

\section{Cryogenic Resonator Experiments}
There are in principle many potential types of GW detectors. In Section 37.3 of the textbook MTW~\cite{MTW}  8 different types of mechanical detectors are discussed, most are not considered as practical.  The expected signals are too small compared to the noise.
To increase the sensitivity of detecting GWs, noise fluctuations must be reduced. For resonator experiments, one must cool down the temperature and increase the quality factor of the resonator. In 1971, Hamilton and Fairbank~\cite{Hamilton:1971} proposed to lower the temperature to 3 mK for a mass of 2.6 ton. In 1974, Braginskii et al.~\cite{Braginskii:1974} remarks on the outlook for increasing the sensitivity of gravitational antennas: ``... for $T = 0.4 \mathrm{K}$ the purely internal losses of the material (sapphire) correspond to a factor of $\omega Q \sim 9 \times 10^{20} \, \mathrm{rad/sec}$. If one could attain this limit for the $\omega Q$ factor, then for $m = 10^4 \, \mathrm{g}, l = 20 \, \mathrm{cm}, T = 0.4 \, \mathrm{K}$ one could record signals of the order of $10^{-3} M_\odot c^2$ at a distance of 1000 Mpc. For $\omega Q = 10^{15} \, \mathrm{rad/sec}$ and the same $T, m, l$, one could expect a response to bursts of gravitational radiation from the nearest galaxies.'' One could see clearly that the mood in the early 1970's is to detect GWs.

The cryogenic resonant bar detectors have already reached a strain spectral sensitivity of $10^{-21} \, \mathrm{Hz^{-1/2}}$ in the kHz region. The Nautilus GW detector had run continuously from December 1995 to December 1996 for a year at a bar temperature of 0.1 K~\cite{Astone:1997gi}. NAUTILUS put an upper limit on periodic sources ranging from $3.4 \times 10^{-23}$ to $1.3 \times 10^{-22}$ depending on frequency in their all-sky search~\cite{Astone:2008up}. The AURIGA-EXPLORER-NAUTILUS-Virgo Collaboration applied a methodology to the search for coincident burst excitations over a 24 h long joint data set~\cite{Acernese:2007ek}. The MiniGRAIL~\cite{Gottardi:2007zn} and Schenberg~\cite{Aguiar:2008zz} cryogenic spherical GW detectors are for omnidirectional GW detection.

\section{Laser Interferometers}
In his history of GW experiments, \textit{Gravity's Shadow}, Collins writes ``The first people to think of using interferometers as a means of detecting gravitational waves seem to have been a pair of Russians, M. Gerstenshtein and V. I. Pustovoit, who published a paper on the idea in Russian in 1962. Independently, Weber and his students considered the idea in 1964. The first person actually to build such a device was Robert Forward, Weber's copioneer, but it was too small to see the waves. Forward, as we can now see, was perhaps the most important member of Weber's team when it came to the constructing the first resonant bar; was the first to analyze a sphere as a detector; and was the first to build an interferometric detector.'' [p 265].\footnote{Moss, Miller and Forward (1971) stated in their Reference 5 that ``To our knowledge, the first suggestion was made by J. Weber in a telephone conversation with one of us (RLF) on 14 September 1964''~\cite{Moss:1971ocz}.}

Experimental work for the first laser interferometer was started in Hughes Research Laboratories in 1966. For a description of the status around that time, let us quote the first paragraph of Section II. A. Wideband Antennas of Moss, Miller and Forward (1971)~\cite{Moss:1971ocz}: ``Since 1966, Hughes Research Laboratories has been engaged in a program to develop wideband gravitational radiation antennas~\cite{Miller:1968}. A long wideband antenna, with a transducer sensitivity equivalent to that obtained on the short resonant antennas, would have a number of advantages over the present detector systems. The increased length would give an increased gravitational radiation capture cross section and increased signal level. (A factor of $10^{10}$ improvement in power sensitivity is possible here.) The broadband characteristic would allow for study of the frequency and phase structure of the radiation signature, giving significant insight into the nature of the source and would allow the phasing of spaced antennas to form a phased array. The broadband characteristic would also allow the use of sophisticated data processing techniques, such as chirp (swept frequency), phase, or comb filters for the extraction of complex signals~\cite{Forward:1967} from the background noise, in addition to the use of standard narrowband frequency filtering for sinusoidal signals, which is the natural filtering action of a resonant antenna.'' In the same paper there is an optical layout of a Michelson interferometer for a gravitational radiation antenna transducer. In this paper, they reported on the achievement of photon-noise-limited performance using $80 \, \mu \mathrm{W}$ from a single mode Spectra-Physics 119 laser in a modified Michelson interferometer on a vibration isolation table in a quiet room. Moss et al. used a piezoelectric driver on one of the interferometer mirrors to generate subangstrom ($3 \times 10^{-14} \, \mathrm{m}$) vibrations of known amplitude (Fig.~\ref{FIG:noiseantenna} [left]). The measured displacement sensitivity of the system in the kilohertz region was $1.3 \times 10^{-14} \, \mathrm{m \, Hz}^{-1/2}$, which compares well with the calculated photon noise limit of $1.06 \times 10^{-14} \, \mathrm{m \, Hz}^{-1/2}$ which was the smallest vibrational displacement measured directly with a laser to that date.

In 1972, Forward and his colleagues completed the first laser-interferometer antenna (Fig.~\ref{FIG:noiseantenna} [right]) for GW detection at the Hughes Research Laboratories, Malibu, California~\cite{Forward:1978zm}: ``The laser interferometer was operated as a detector for gravitational radiation for 150 h during the nights and weekends from the period 4 October through 3 December 1972. During the same period, bar antennas were operated by the Maryland, Glasgow, and Frascati groups, with 18 events reported by the Frascati group in their single bar, 22 single bar events and no coincidences reported by the Glasgow group in their two bars, and 28 coincidences reported by the Maryland group between the Argonne bar and the Maryland bar and/or disk antennas. The various bar antenna systems were quite different but in general were sensitive to gravitational-radiation strain spectral components with an amplitude of the order of 0.1 fm/m in a narrow band of frequencies about the resonant frequency of the bar. The wideband interferometer data was analyzed by ear, with the detection sensitivity estimated to be of the order of 1 -- 10 fm/m (depending upon the signature of the signal) for the total of the gravitational-radiation strain spectral components in the band from 1 -- 20 kHz. No significant correlations between the Malibu interferometer output and any of the bar events or coincidences were observed. ... Thus, at the time one of the bar-antenna systems produced an event or coincidence corresponding to a gravitational-radiation signal with an amplitude of 0.1 fm/m due to spectral components in a narrow band around the bar resonance, the amplitude of the gravitational-radiation spectral components in the entire band from 1 -- 20 kHz was definitely less than 10 fm/m and was probably less than 1 fm/m.''

\begin{figure}[ht]
\begin{center}
\includegraphics[width=2.5in]{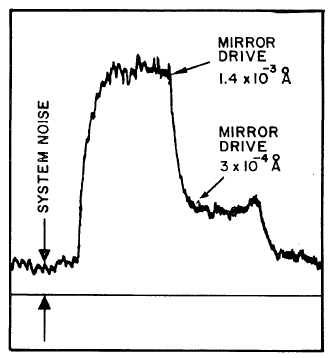}
\hspace{0.5cm}
\includegraphics[width=2.5in]{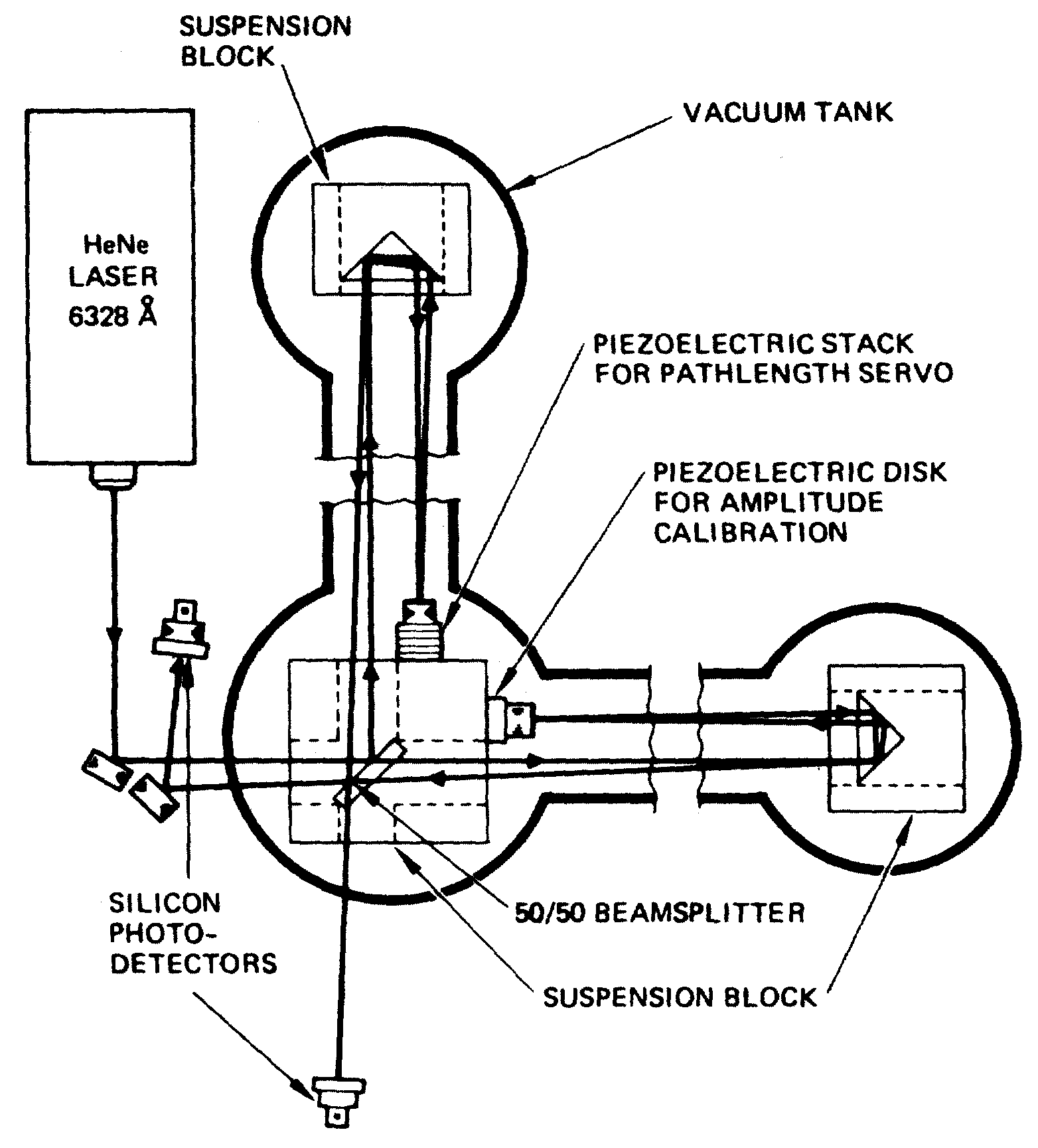}
\caption{(left) Interferometer system noise measurement at 5 kHz of Moss, Miller and Forward (1971)~\cite{Moss:1971ocz}; (right) Schematic of Malibu Laser Interferometer GW Antenna (from Forward 1978)~\cite{Forward:1978zm}.} \label{FIG:noiseantenna}
\end{center}
\end{figure}

In the comprehensive textbook MTW~\cite{MTW} on page 1014 one can read.
``...or by laser interferometry. Several features of such detectors are explored in exercises 37.6 and 37.7. As shown in exercise 37.7, such detectors have so low a sensitivity that they are of little experimental interest.''  This was a reasonable assessment based on the technology in 1972, when that book was written.

Collins writes ``It is Rainer Weiss of MIT who is now almost universally credited with the conceptualization of an interferometer capable of seeing the kind of flux of gravitational waves the theorists believed might be there to be seen. ...Weiss himself acknowledges Felix Pirani and Phillip Chapman (see below), and we also have the Russians as precursors of the idea.''~\cite{Collins04}, p 266.

In 1972, a classical source document appeared for large-scale laser interferometry for GW detection in the Quarterly Progress Report, Research Laboratory of Electronics (RLE) of MIT. The gravitation research section contained the report of the Rai (Rainer) Weiss group [consisting of academic and research staff---Prof. R. Weiss, Dr. D. J. Muehner, R. L. Benford, and graduate students---D. K. Owens, N. A. Pierre, M. Rosenbluh]; Subsection A is on ``Balloon measurements of far infrared background radiation'' with the names D. J. Muehler and R. Weiss at the end of this subsection; Subsection B is on ``electromagnetically coupled broadband gravitational antenna'' with the name R. Weiss at the end of this subsection. Subsection B is the classical source document commonly called the Weiss RLE report~\cite{Weiss:1972}. It contained the motivation and the proposed antenna design together with a rather thorough discussion of the most fundamental noise sources in the interferometer. The fundamental noise sources discussed include:

a. Amplitude noise in the laser output power,

b. Laser phase noise or frequency instability,

c. Mechanical thermal noise in the antenna,

d. Radiation-pressure noise from laser light,

e. Seismic noise,

f. Thermal-gradient noise,

g. Cosmic-ray noise,

h. Gravitational-gradient noise,

i. Electric field and magnetic field noise.

In the same year Weiss submitted a proposal to the National Science Foundation but failed to obtain funding. Weiss submitted a proposal again in 1974 for a January startup. The proposal was sent to Peter Kafka of Max Planck Institute at Munich for review. Kafka passed it to his experimental colleagues for advice and was shocked when his colleagues took it up. (The German group had a resonant GW detector at that time.) Weiss did not receive the proposal acceptance letter from NSF until May 1975. German interests might have been a favorable factor for final NSF approval.

Braginskii found there was an issue of quantum-mechanical limit on the sensitivities of GW detectors and advocated Quantum Nondemolition (QND) measurement. Braginskii showed that interferometers could avoid the quantum issue more easily than resonant detectors. It might be because of this issue that Drever had decided to switch his main efforts to interferometry in 1976 at an Erice conference, [see Collins~\cite{Collins04}, p~285].

By 1980 Thorne (following discussions with Weiss) changed his mind and had become an interferometer advocate. In the same year, Caves found a scheme which used the squeezed vacuum to alleviate the shot noise of an interferometer. Although the cryogenic resonator GW detectors would still take a decade to flourish, the interferometer GW detector approach was set to become the main approach. It is about this time, in 1981 Faller and Bender presented the first public proposal of a space interferometer for GW detection. In 1986--1987, construction was started on the ISAS (The Institute of Space and Aeronautical Science) Tenko 10 m Interferometer and Physics Department 3 m Interferometer at Tokyo University. A list of laser interferometers with independently suspended mirrors is given in the Table~\ref{TABLE:LaserMirror}. In the table, we also list the interferometers with independently suspended mirrors for the QED vacuum birefringence experiment Q\&A. It measures the birefringence of the vacuum due to magnetic field modulation.

Major detection efforts in the high frequency band are in the long arm laser interferometers. The TAMA 300 m arm length interferometer~\cite{TAMA300}, the GEO 600 m interferometer~\cite{GEO600} and the kilometer size laser-interferometric GW detectors --- LIGO~\cite{LIGO} (two 4 km arm length, one 2 km arm length) and VIRGO~\cite{VIRGO}; all achieved their original sensitivity goals basically. Around the frequency 100 Hz, the LIGO and Virgo sensitivities are both at the level of $10^{-23} \, \mathrm{Hz}^{-1/2}$.

These interferometers paved the road for the second generation detectors - adLIGO, adVirgo and KAGRA. The discovery of adLIGO along with more discussion, including some on the third generation detectors will be presented in sections~\ref{DirectDetection} and~\ref{outlook}.

\begin{table}\
\begin{center}
\begin{tabular}{|c|c|c|c|}
\hline
Interferometer & Arm  & Effective Optical & Year Construction \\
 & Length [m] & Path Length [km] & Started \\
\hline\hline
Hughes Research Lab (HRL)~\cite{Miller:1968, Moss:1971ocz, Forward:1978zm} & 2 & 0.0085 (N=4) & 1966 \\
\hline
MIT prototype~\cite{Weiss:1972} & 1.5 & 0.075 (N=50) & 1971 \\
\hline
Garching 3 m prototype & 3 & 0.012 (N=4) & 1975 \\
\hline
Glasgow 1 m prototype~\cite{email} & 1 & 0.036 (N=36; in static test & 1976 \\
 &  & reached N=280) & \\
\hline
Glasgow 10 m prototype~\cite{email} & 10 & 25.5 (F-P: F=4000) & 1980 \\
\hline
Caltech 40 m prototype & 40 & 75 & 1980 \\
\hline
Garching 30 m prototype & 30 & 2.7 (N=90) & 1983 \\
\hline
ISAS Tenko 10 m prototype~\cite{Kawamura1989} & 10 & 1 (N=100) & 1986 \\
\hline
U. Tokyo prototype~\cite{Kawabe1996, Ando:1998dq} & 3 & 0.42 (F-P: F=220) & 1987 \\
\hline
ISAS Tenko 100 m prototype~\cite{Kawashima1991, Miyoki1997a, Miyoki1997b, Miyoki1997c} & 100 & 10 (N=100) & 1991 \\	
\hline
NAOJ 20 m prototype~\cite{Araya1997} & 20 & 4.5 (F-P: F=350) & 1991 \\	
\hline
Q\&A 3.5 m prototype~\cite{Chen:2006cd} & 3.5 & 67 (F-P: F=30000) & 1993 \\
\hline
TAMA 300 m~\cite{TAMA300} & 300 & 96 (F-P: F=500) & 1995 \\
\hline
GEO	600 m~\cite{GEO600, wiki} & 600 & 1.2 (N=2) & 1995 \\
\hline
LIGO Hanford (2 km)~\cite{LIGO, Abbott:2007kv} & 2000 & 143 (F-P: F=112) & 1994 \\
\hline
LIGO Hanford (4 km)~\cite{LIGO, Martynov:2016fzi} & 4000 & 1150 (F-P: F=450) & 1994 \\
\hline
LIGO Livingston (4 km)~\cite{LIGO, Martynov:2016fzi} & 4000 & 1150 (F-P: F=450) & 1995 \\
\hline
VIRGO~\cite{VIRGO, TheVirgo:2014hva} & 3000 & 850 (F-P: F=440) & 1996 \\
\hline
AIGO prototype~\cite{Zhao:2015cna, Zhao2011} & 80 & 760/66 (F-P: east arm F=15000; & 1997 \\
 & & south arm F=1300) &  \\
\hline
LISM~\cite{Sato:2004sz} & 20 & 320 (F-P: F=25000) & 1999 \\
\hline
CLIO 100 m cryogenic~\cite{Agatsuma:2009tm} & 100 & 190 (F-P: F=3000) & 2000 \\
\hline
Q\&A 7 m~\cite{Mei:2010aq} & 7 & 450 (F-P: F=100000) & 2008 \\
\hline
LCGT/KAGRA~\cite{KAGRA, Aso:2013eba} & 3000 & 2850 (F-P: F=1500) & 2010 \\
\hline
Q\&A 9 m~\cite{QA9m} & 9 & 570 (F-P: F=100000) & 2016 \\
\hline
LIGO India~\cite{IndIGO} & 4000 & 1150 (F-P: F=450) & 2016 \\
\hline
ET~\cite{Hild:2008ng} & 10000 & 3200 (F-P: F$\sim$500) & proposal under study \\
\hline
\end{tabular}
\end{center}
\caption{Laser interferometers with independently suspended mirrors. In third column, in the parenthesis either the number N of paths is given or Fabry-Perot Finesse F is given.} \label{TABLE:LaserMirror}
\end{table}

\subsection{Space interferometers}
Doppler tracking of spacecraft is also interferometry and can be used to constrain (or detect) the level of low-frequency GWs~\cite{Estabrook:1975}. The separated test masses of this GW detector are the Doppler tracking radio antenna on Earth and a distant spacecraft. Doppler tracking measures relative distance change. Estabrook and Walquist analyzed~\cite{Estabrook:1975} the effect of GWs passing through the line of sight of spacecraft on the Doppler tracking frequency measurements (see also Ref.~\cite{Wahlquist:1987rx}). From these measurements, GWs can be detected or constrained. The most recent measurements came from the Cassini spacecraft Doppler tracking (CSDT). Armstrong, Iess, Tortora, and Bertotti~\cite{Armstrong:2003ay} used precision Doppler tracking of the Cassini spacecraft during its 2001--2002 solar opposition to derive improved observational limits on an isotropic background of low-frequency gravitational waves. They used the Cassini multilink radio system and an advanced tropospheric calibration system to remove the effects of the leading noises --- plasma and tropospheric scintillation --- to a level below the other noises. The resulting data were used to construct upper limits on the strength of an isotropic background in the $1 \, \mu \mathrm{Hz}$ to $1 \mathrm{mHz}$ band~\cite{Armstrong:2003ay}. The characteristic strain upper limit curve labeled CSDT in Fig.~\ref{FIG:strain} is a smoothed version of the curve in the Fig.~4 of Ref.~\cite{Armstrong:2003ay}. The corresponding CSDT curves on the strain psd amplitude in Fig.~\ref{FIG:PSDamplitude} and the normalized spectral energy density in Fig.~\ref{FIG:energydensity} are calculated using Table~\ref{TABLE:conversion} for conversion.

Space laser interferometers for GW detection (eLISA/LISA~\cite{Jennrich:2011, LISA:2000}, ASTROD~\cite{Ni:2008zz, Bec:2000}, ASTROD-GW~\cite{Ni:2009aa, Ni:2009bb, Ni:2010ur, Ni:2010nr, Ni:2012eh, Ni:2016}, ASTROD-EM~\cite{Ni:2016, NiWu}, Super-ASTROD~\cite{Ni:2008bj}, DECIGO~\cite{Kawamura:2006up}, Big Bang Observer~\cite{Crowder:2005nr}, ALIA~\cite{Bender:2004vw}, ALIA-descope (Taiji)~\cite{Gong:2014mca}, gLISA (GEOGRAWI)~\cite{deAraujo:2011aa, Tinto:2011nr, Tinto:2014eua}, GADFLI~\cite{McWilliams:2011ra}, LAGRANGE~\cite{Conklin:2011vt}, OMEGA~\cite{Hellings:2011} and TIANQIN~\cite{Luo:2015ght}) hold the most promise with high signal-to-noise ratio. LISA~\cite{LISA:2000} (Laser Interferometer Space Antenna) is aimed at detection of $10^{-4}$ to $1 \, \mathrm{Hz}$ GWs with a strain sensitivity of $4 \times 10^{-21}/\mathrm{Hz}^{1/2}$ at $1 \, \mathrm{mHz}$, see Table~\ref{TABLE:GWmission}. There are abundant sources for eLISA/LISA, ASTROD, ASTROD-GW and Earth-orbiting missions: (i) In our Galaxy: galactic binaries (neutron stars, white dwarfs, etc.); (ii) Extra-galactic targets: supermassive black hole binaries, supermassive black hole formation; and (iii) Cosmic GW background. A date of launch of eLISA or a substitute mission is set around 2034~\cite{EAS:2013}.

LISA Pathfinder was launched on December 3, 2015 and successfully tested the drag-free technology to satisfy not just the requirement of LISA Pathfinder, but also the requirement of LISA~\cite{Armano:2016bkm}. The success paved the road for all the space mission proposals in the table. And we do anticipate the possibility of an earlier launch date for eLISA (or a substitute mission) and possibly an earlier flight for other missions.

\begin{table}[ht]
\begin{center}
\begin{tabular}{|c|c|c|c|c|}
\hline
Mission Concept & S/C Configuration & Arm & Orbit & S/C \\
 & & Length & Period & \# \\
\hline
\multicolumn{5}{|c|}{\it Solar-Orbit GW Mission Proposals} \\
\hline
LISA~\cite{LISA:2000} & Earth-like solar orbits with $20^\circ$ lag & 5 Gm & 1 year & 3 \\
\hline
eLISA~\cite{Jennrich:2011} & Earth-like solar orbits with $10^\circ$ lag & 1 Gm & 1 year & 3 \\
\hline
ASTROD-GW~\cite{Ni:2012eh} & Near Sun-Earth L3, L4, L5 points & 260 Gm & 1 year & 3 \\
\hline
Big Bang Observer~\cite{Crowder:2005nr} & Earth-like solar orbits & 0.05 Gm	& 1 year & 12 \\
\hline
DECIGO~\cite{Kawamura:2006up} & Earth-like solar orbits & 0.001 Gm & 1 year & 12 \\
\hline
ALIA~\cite{Bender:2004vw} & Earth-like solar orbits & 0.5 Gm & 1 year & 3 \\
\hline
Taiji~\cite{Gong:2014mca} & Earth-like solar orbits & 3 Gm & 1 year & 3 \\
\hline
Super-ASTROD~\cite{Ni:2008bj} & Near Sun-Jupiter L3,\! L4,\! L5 points (3 S/C), & 1300 Gm & 11 year & 4 or 5 \\
 & Jupiter-like solar orbit(s)(1-2 S/C) & & & \\
\hline
\multicolumn{5}{|c|}{\it Earth-Orbit GW Mission Proposals} \\
\hline
OMEGA~\cite{Hellings:2011} & 0.6 Gm height orbit & 1 Gm & 53.2 days & 6 \\
\hline
gLISA/GEOGRAWI~\cite{deAraujo:2011aa, Tinto:2011nr, Tinto:2014eua} & Geostationary orbit & 0.073 Gm & 24 hours & 3 \\
\hline
GADFLI~\cite{McWilliams:2011ra} & Geostationary orbit & 0.073 Gm & 24 hours & 3 \\
\hline
TIANQIN~\cite{Luo:2015ght} & 0.057 Gm height orbit & 0.11 Gm & 44 hours & 3 \\
\hline
ASTROD-EM~\cite{Ni:2016, NiWu} & Near Earth-Moon L3, L4, L5 points & 0.66 Gm & 27.3 days & 3 \\
\hline
LAGRANGE~\cite{Conklin:2011vt} & Near Earth-Moon L3, L4, L5 points & 0.66 Gm & 27.3 days & 3 \\
\hline
\end{tabular}
\end{center}
\caption{A Compilation of GW Mission Proposals} \label{TABLE:GWmission}
\end{table}

\section{Pulsar Timing Arrays}
Now there are 4 major pulsar timing arrays (PTAs): the European PTA (EPTA)~\cite{EPTA}, the NANOGrav~\cite{NANOGrav}, the Parks PTA (PPTA)~\cite{PPTA} and the International (EPTA, NANOGrav and PPTA combined) PTA (IPTA)~\cite{IPTA}. For recent reviews on pulsar timing and pulsar timing arrays for GW detection, please see~\cite{Manchester:2015, Zhu:2015ara}. These 4 PTAs have improved greatly on the sensitivity for GW detection recently~\cite{Lentati:2015qwp, Shannon:2015ect, Arzoumanian:2015liz}. Upper limits on the isotropic stochastic background from EPTA, PPTA and NANOGrav are listed in Table~\ref{TABLE:Ulimit}.

The most stringent limit is from Shannon et al.~\cite{Shannon:2015ect} using observations of millisecond pulsars from the Parks telescope to constrain $A_\mathrm{yr}$ to less than $1.0 \times 10^{-15}$ with 95\% confidence. This limit already excludes the present and most recent model predictions with 91 -- 99.7\% probability~\cite{Shannon:2015ect}. The three experiments form a robust upper limit of $1 \times 10^{-15}$ on $A_\mathrm{yr}$ at 95\% confidence level ruling out most models of supermassive black hole formation.

\begin{table}[ht]
\begin{center}
\begin{tabular}{|c||c|c|c|c|}
\hline
 & No. of & No. of & Observation & Constraint on characteristic strain  \\
 & pulsars & years & radio band & $h_c(f) [= A_\mathrm{yr} (f/\mathrm{1 yr^{-1}})^{-2/3},$ \\
 & included & observed & [MHz] & $(f = 10^{-9}$ -- $10^{-7} \mathrm{Hz}$)] \\
\hline\hline
 EPTA~\cite{EPTA} & 6 & 18 & 120 -- 3000 & $A_\mathrm{yr} < 3 \times 10^{-15}$ \\
\hline
 PPTA~\cite{PPTA} & 4 & 11 & 3100 & $A_\mathrm{yr} < 1 \times 10^{-15}$ \\
\hline
 NANOGrav~\cite{NANOGrav} & 27 & 9 & 327 -- 2100 & $A_\mathrm{yr} < 1.5 \times 10^{-15}$ \\
\hline
\end{tabular}
\end{center}
\caption{Upper limits on the isotropic stochastic background from 3 pulsar timing arrays.} \label{TABLE:Ulimit}
\end{table}

\section{Classification of GWs and Other GW Experiments}

The frequency classification for gravitational waves is presented in Table~\ref{TABLE:freqGW}.  Figure~\ref{FIG:strain} shows the strain vs frequency for various detectors and sources.  Figure~\ref{FIG:PSDamplitude} displays the strain power spectral density (psd) amplitude vs.\ frequency for various GW detectors and GW sources. Figure~\ref{FIG:energydensity} presents the Normalized GW spectral energy density $\Omega_\mathrm{gw}$ vs.\ frequency for the GW detector sensitivities and GW sources.
For more details on other GW experiments, please see~\cite{Kuroda:2015owv} and the references therein.

\begin{table}[ht]
\begin{center}
\begin{tabular}{|p{6cm}|p{9cm}|}
\hline
Frequency band & Detection method \\
\hline\hline
Ultrahigh frequency band: \newline above 1 THz  & Terahertz resonators, optical resonators, and magnetic conversion detectors \\
\hline
Very high frequency band: \newline 100 kHz -- 1 THz  & Microwave resonator/wave guide detectors, laser \newline interferometers and Gaussian beam detectors \\
\hline
High frequency band (audio band)$^*$: \newline 10 Hz -- 100 kHz  & Low-temperature resonators and ground-based laser-interferometric detectors \\
\hline
Middle frequency band: \newline 0.1 Hz -- 10 Hz  & Space laser-interferometric detectors of arm length 1,000 km -- 60,000 km \\
\hline
Low frequency band (milli-Hz \newline band)$^\dagger$: 100 nHz -- 0.1 Hz  & Space laser-interferometric detectors of arm length longer than 60,000 km \\
\hline
Very low frequency band (nano-Hz \newline band): 300 pHz -- 100 nHz  & Pulsar timing arrays (PTAs) \\
\hline
Ultralow frequency band: \newline 10 fHz -- 300 pHz  & Astrometry of quasar proper motions \\
\hline
Extremely low (Hubble) frequency band (cosmological band): \newline 1 aHz -- 10 fHz  & Cosmic microwave background experiments \\
\hline
Beyond Hubble-frequency band: \newline below 1 aHz  & Through the verifications of inflationary/primordial cosmological models \\
\hline
\multicolumn{2}{l}{${}^*$The range of audio band normally goes only to 10 kHz.} \\
\multicolumn{2}{l}{${}^\dagger$The range of milli-Hz band is 0.1 mHz to 100 mHz.}
\end{tabular}
\end{center}
\caption{Frequency Classification of Gravitational Waves.} \label{TABLE:freqGW}
\end{table}

\begin{figure}[ht]
\begin{center}
\includegraphics[width=6in]{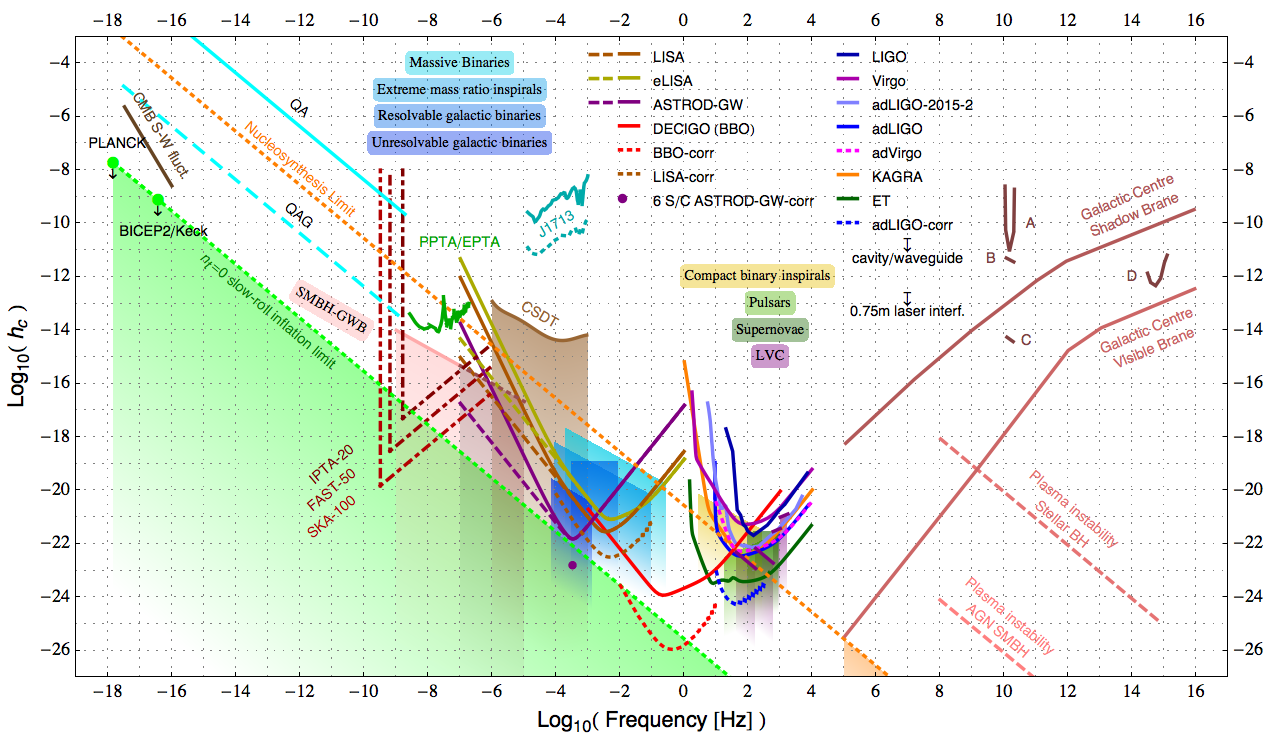}
\caption{Characteristic strain $h_c$ vs. frequency for various GW detectors and sources (from~\cite{Kuroda:2015owv}). [QA: Quasar Astrometry; QAG: Quasar Astrometry Goal; LVC: LIGO-Virgo Constraints; CSDT: Cassini Spacecraft Doppler Tracking; SMBH-GWB: Supermassive Black Hole-GW Background.]} \label{FIG:strain}
\end{center}
\end{figure}

\begin{figure}[ht]
\begin{center}
\includegraphics[width=6in]{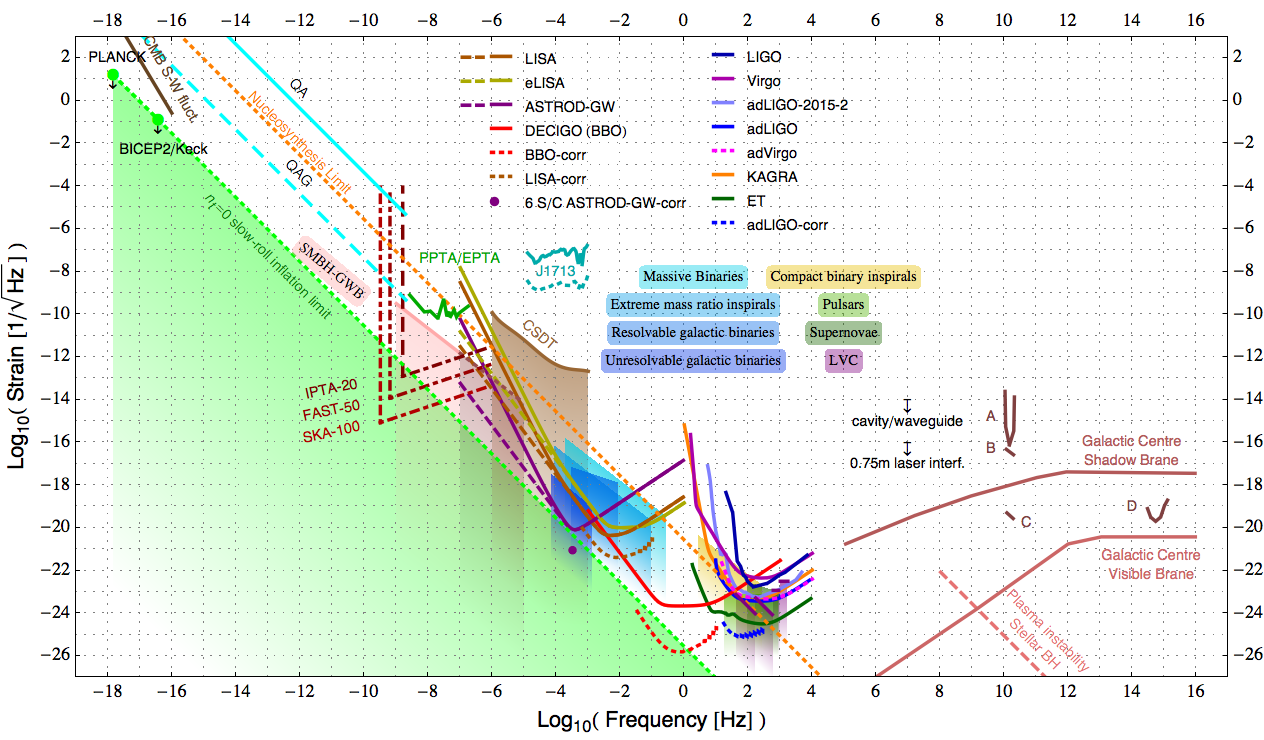}
\caption{Strain power spectral density (psd) amplitude vs. frequency for various GW detectors and GW sources (from~\cite{Kuroda:2015owv}). See Fig.~\ref{FIG:strain} caption for the meaning of various acronyms.} \label{FIG:PSDamplitude}
\end{center}
\end{figure}

\begin{figure}[ht]
\begin{center}
\includegraphics[width=6in]{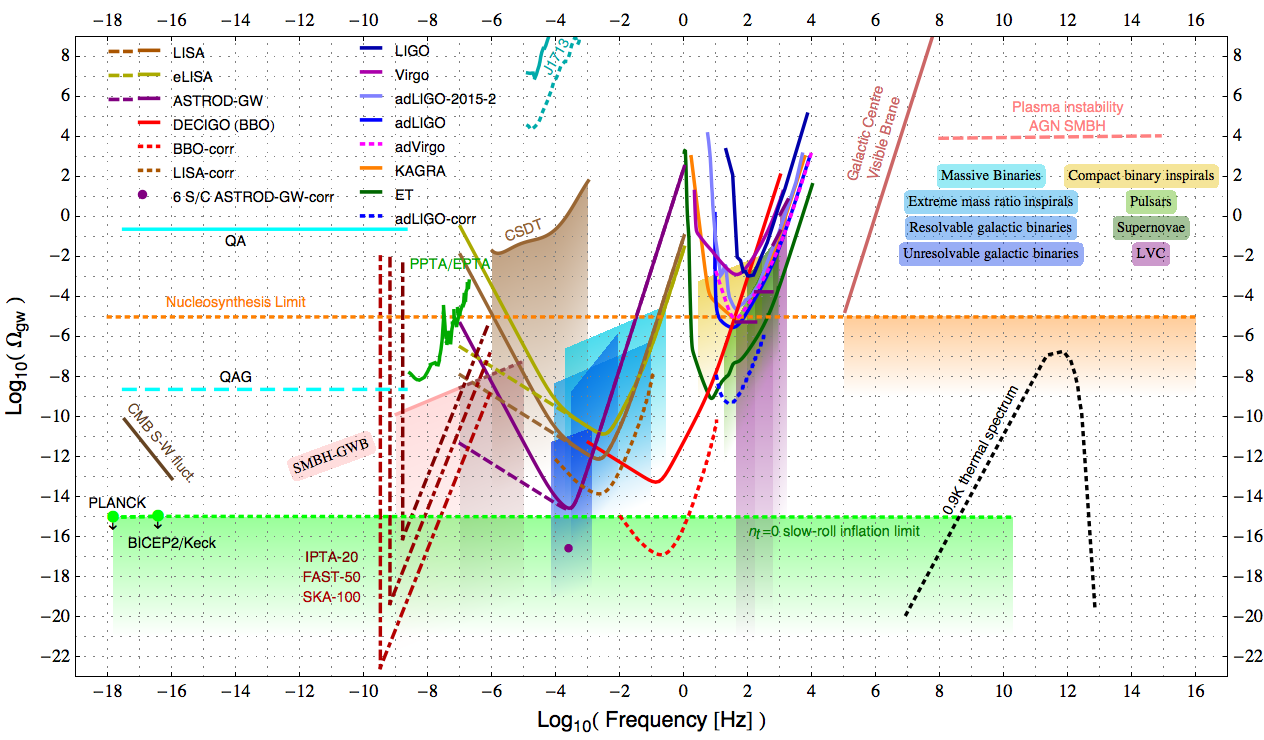}
\caption{Normalized GW spectral energy density $\Omega_\mathrm{gw}$ vs. frequency for GW detector sensitivities and GW sources (from~\cite{Kuroda:2015owv}). See Fig.~\ref{FIG:strain} caption for the meaning of various acronyms.} \label{FIG:energydensity}
\end{center}
\end{figure}

\section{Numerical Relativity}
Numerical relativity (NR) is an essential compliment to the theoretical and experimental efforts.  The Einstein equations are a rather complex non-linear system of dynamical and constraint equations.  Analytic approaches alone are not able to  predict with sufficient accuracy the  waveforms that can be expected from the various astrophysical sources.

From the early days it was clear that numerical calculations would be needed.
Some of the main issues are (i) how to choose the coordinate gauge conditions, (ii) how to achieve adequate numerical stability, (iii) how to set up the evolution equations to maintain the constraint preservation, and (iv)  how to deal with the central singularity (two methods, puncture and excision).   A main objective is to model binary coalesecence through the inspiral, merger and ringdown phases. It took many years of improving the computational power and techniques until it was possible to successfully model the coalescence of two binaries.  After much effort a large number of templates have been produced that include a variety of expected waveforms from binary inspiral that can be, and have been, used with matched filtering to recognize a small GW signal within the noise.

Reference~\cite{Cardoso:2014uka} has a table including the milestones in the progress of numerical relativity.  Here we have adapted some excerpts related to our concerns.

\begin{small}
\begin{itemize}

\item[1964] First documented attempt to solve Einstein's equations numerically by Hahn \& Lindquist~\cite{385}.

\item[1966]  May and White perform a full nonlinear numerical collapse simulation for some realistic
equations of state~\cite{543}.

\item[1970] Vishveshwara~\cite{762} studies numerically the scattering of GWs by BHs: finding at late times ringdown waves: quasi-normal modes (QNM).

\item[1975] The NR theses of Smarr \& Eppley~\cite{311, 710}.

\item[1977] NR is born with coordinated efforts to evolve BH spacetimes~\cite{708, 287, 711}.

\item[1992] Bona and Mass{\'o} show that harmonic slicing has a singularity-avoidance property, setting the stage for the development of the 1+log slicing~\cite{115}.

\item[1993] First successful simulation of the head-on collision of two BHs, QNM ringing of the final BH observed~\cite{37}.

\item[1993] Choptuik uses mesh refinement and finds evidence of universality and scaling in the gravitational collapse of a massless scalar field~\cite{212}.

\item[1994] The ``Binary Black Hole Grand Challenge Project'', the first large collaboration with the aim of solving a specific NR problem (modeling a binary BH coalescence), is launched~\cite{542, 213}.

\item[1995]  Through a conformal decomposition, a split of the extrinsic curvature and use of additional variables Shibata \& Nakamura~\cite{695} and  Baumgarte \& Shapiro~(1998)~\cite{78} recast the ADM~\cite{ADM} Hamiltonian equations as the so-called BSSN system.

\item[1996] Br{\"u}gmann~\cite{140} uses mesh refinement for simulations of BH spacetimes.

\item[1998] First stable simulations of a single BH spacetime in fully 4 dimensional NR within a ``characteristic formulation''~\cite{508, 362}, and two years later within a Cauchy formulation~\cite{23}.

\item[2000] The first general relativistic simulation of the merger of two NSs~\cite{698}.

\item[2005] Pretorius~\cite{629} achieves the first long-term stable numerical evolution of a BH binary.

\item[2006] Soon afterwards, other groups independently succeed in evolving merging BH binaries using different
techniques~\cite{159, 65}.

\item[2006] NR simulations of BH-NS binaries~\cite{699}.

\item[2007] Unprecedented accuracy and number of orbits achieved in simulating a BH binary through inspiral and merger with a spectral code that later becomes known as ``SpEC''~\cite{122} and uses multi-domain decomposition~\cite{618} and a dual coordinate frame~\cite{678}.

\end{itemize}
\end{small}

For more details regarding NR the reader may consult some textbooks and additional papers, e.g.~\cite{Al08, BS10, 509, SLS09, Yo:2015kta}.

\section{Direct detection} \label{DirectDetection}
In 2007 Kennefick wrote ``Today it is widely expected that the first direct detection of gravitational waves will take place within the decade''~\cite{Kennefick} p 1.  Pretty accurate.

From the LIGO O1 observation run, there are two GW events with significance greater than 5.3-$\sigma$ and one likely GW candidate with significance 1.7-$\sigma$ in 130 days of observation~\cite{Abbott:2016blz, Abbott:2016nmj}. These detection events/candidate occurred on September 14, October 12 and December 26 of 2015 with luminosity distances of about 420 Mpc, 1000 Mpc and 440 Mpc, respectively. Their radiated energies are the equivalent energies of 3, 1.0 and 1.5 solar mass respectively. All events had a peak luminosity around $3 \times 10^{56} \, \mathrm{erg \, s^{-1}}$. The peak intensity of the first detected event GW150914 reached a strain of $10^{-21}$ at a frequency around 100 Hz. These values together with some other characteristics deduced from the LIGO O1 GW observations are listed in Table~\ref{TABLE:2+1GW}.

The O1 observation period spanned 130 days from September 12, 2015 to January 19, 2016. If we simply scale this to one year of observation, we would have a chance to observe five GW events with significance greater than 5.3-$\sigma$. When the advanced LIGO goal sensitivity is reached, there would be a 3 fold improvement and a 27 fold reach in volume (still basically the local universe); hence more than 100 GW events per year with significance greater than 5.3-$\sigma$. For a third-generation earth-based laser interferometer for GW detection (like ET), there would be another tenfold increase in sensitivity and 1000-fold reach in volume; hence about 100 k GW events per year.
This would be a lot of new information for astronomical and cosmological studies.

\begin{table}[ht]
\begin{center}
\begin{tabular}{|c||c|c|c|}
\hline
Event & \qquad GW150914 \qquad{} & \qquad GW151226 \qquad{} & \qquad LVT151012 \qquad{} \\
\hline\hline
Signal-to-noise ratio $\rho$ & $23.7$ & $13.0$ & $9.7$ \\
\hline
Significance & $> 5.3$-$\sigma$ & $> 5.3$-$\sigma$ & $1.7$-$\sigma$ \\
\hline
Primary mass $M^\mathrm{source}_1/M_\odot$ & $36.2^{+5.2}_{-3.8}$ & $14.2^{+8.3}_{-3.7}$ & $23^{+18}_{-6}$ \\
\hline
Secondary mass $M^\mathrm{source}_2/M_\odot$ & $29.1^{+3.7}_{-4.4}$ & $7.5^{+2.3}_{-2.3}$ & $13^{+4}_{-5}$ \\
\hline
Effective inspiral spin $\chi_\mathrm{eff}$ & $-0.06^{+0.14}_{-0.14}$ & $0.21^{+0.20}_{-0.10}$ & $0.0^{+0.3}_{-0.2}$ \\
\hline
Final mass $M^\mathrm{source}_f/M_\odot$ & $62.3^{+3.7}_{-3.1}$ & $20.8^{+6.1}_{-1.7}$ & $35^{+14}_{-4}$ \\
\hline
Final spin $a_f$ & $0.68^{+0.05}_{-0.06}$ & $0.74^{+0.06}_{-0.06}$ & $0.66^{+0.09}_{-0.10}$ \\
\hline
Radiated energy $E_\mathrm{rad}/(M_\odot c^2)$ & $3.0^{+0.5}_{-0.4}$ & $1.0^{+0.1}_{-0.2}$ & $1.5^{+0.3}_{-0.4}$ \\
\hline
Peak luminosity $l_\mathrm{peak}/(\mathrm{erg \, s^{-1}})$ & $3.6^{+0.5}_{-0.4} \times 10^{56}$ & $3.3^{+0.8}_{-1.6} \times 10^{56}$ & $3.1^{+0.8}_{-1.8} \times 10^{56}$ \\
\hline
Luminosity distance $D_L/\mathrm{Mpc}$ & $420^{+150}_{-180}$ & $440^{+180}_{-190}$ & $1000^{+500}_{-500}$ \\
\hline
Source redshift $z$ & $0.09^{+0.03}_{-0.04}$ & $0.09^{+0.03}_{-0.04}$ & $0.2^{+0.09}_{-0.09}$ \\
\hline
\end{tabular}
\end{center}
\caption{Characteristics of 2 GW events and one GW candidate deduced from LIGO O1 GW observations~\cite{Abbott:2016blz, Abbott:2016nmj}.} \label{TABLE:2+1GW}
\end{table}

\section{Outlook} \label{outlook}
With the LIGO discovery announcements of February and June, two important things are verified: (i) GWs are directly detected in the solar-system; (ii) Black holes (BHs), binary BHs and BH coalescences are discovered and measured experimentally and directly with the distances reached more than 1 Gly. These observations have already given some constraints on the mass of massive gravity and on the speed of the GWs. More observations would test the fundamental laws of gravity to a much greater extent. With the sensitivity goal reached for second-generation Earth-based interferometers for GW detection, the expected GW events at more than 5-$\sigma$ confidence level will be more than 100 per year and positions will be moderately determined. With the third-generation interferometers constructed and completed, the expected GW events at more than 5-$\sigma$ confidence level will be more than 100,000 per year; those at more than 50-$\sigma$ confidence level will be more than 100 per year with their 3-d position determined to 3\% in distance and to sub-degree in angular position.

With LISA Pathfinder having successfully tested the drag-free technology to satisfy not just the requirement of LISA Pathfinder, but also~\cite{Armano:2016bkm} basically the requirement of LISA along with the first detection of GWs by aLIGO, we do anticipate the possibility of an earlier launch date for eLISA (or a substitute mission) and likely an earlier flight of other missions. Although the prospect of a launch of space GW missions is only expected a decade later, the detection in the low frequency band may have the largest signal to noise ratios. This will enable the detailed study of black hole co-evolution with galaxies and with the dark energy issue. Foreground separation and correlation detection method needs to be investigated to achieve the sensitivities $10^{-16}$ -- $10^{-17}$ or beyond in $\Omega_\mathrm{gw}$ to study the primordial GW background for exploring the very early universe and possibly quantum gravity regimes.

Another avenue for real-time direct detection is from the PTAs. The PTA bound on the stochastic GW background already excludes most theoretical models; this may mean we could detect very low frequency GWs anytime too with a longer time scale. With the current technology development and astrophysical understanding, we are in a position to use GWs to study more thoroughly galaxies, supermassive black holes and clusters together with cosmology, and to explore deeper into the origin of gravitation and our universe. The next 100 years will be the golden age of GW astronomy and GW physics.

\section{Conclusion}
Here we have presented a brief history of gravitational waves in the first century of GR: the controversies during the long theoretical development, the experiments aimed at direct detection, the impact of the discovery of the binary pulsar, the quite recent direct interferometer observations, and the proposals and plans for future efforts. With a glimpse of future GW sensitivities in Figs.~\ref{FIG:strain} -- \ref{FIG:energydensity}, prospects for the next century look even more promising for the current and coming generations.

We can think of no more fitting way to end than with some words from William Blake, also quoted at the end of both MTW~\cite{MTW} and Kennefick~\cite{Kennefick}, which seems to us to be now even more apt:
``What is now proved was once only imagin'd''~\cite{Blake}.

\section*{Addendum}
After our review was submitted we learned that a nice, readable article with some similarity had just appeared~\cite{Jorge L. Cervantes-Cota:2016ewh}. There is a lot of overlap in the topics considered in these two works, while in other ways they are complimentary. We encourage our readers to also look at that work.

\section*{Acknowledgements}
We would like to thank H.-J.~Yo for his helpful thoughts on the topic of numerical relativity.
We also thank M.~Ando, J.~Hough, L.~Ju, S.~Kawamura, S.~Miyoki, K.~Somiya and C.~Zhao for their prompt e-mail replies on questions about finesses of various interferometers.
CMC was supported by the Ministry of Science and Technology of the ROC under the grants MOST 102-2112-M-008-015-MY3 and 105-2112-M-008-018.
WTN would like to thank Science and Technology Commission of Shanghai Municipality (STCSM-14140502500) and Ministry of Science and Technology of China (MOST-2013YQ150829, MOST-2016YFF0101900) for supporting this work in part.


\end{document}